\documentclass[prd,twocolumn,showpacs,showkeys,preprintnumbers,amsmath,amssymb]{revtex4}

\usepackage{graphicx}
\usepackage{dcolumn}
\usepackage{bm}

\begin{document}

\title{Strong Field Gravitational Lensing by a Kerr Black Hole}

\author{S. E. V\'azquez}
\email{svazquez@physics.ucsb.edu}
 \affiliation{Department of Physics, University of California \\Santa Barbara, California 93106-9530}
\author{E. P. Esteban}
 \email{ep_esteban@webmail.upr.clu}
\affiliation{ Department of Physics and Astronomy MS108 Rice
University 6100 Main Street \\  Houston, TX 77005-1892}
\affiliation{Department of Physics, University of Puerto Rico \\
Humacao, Puerto Rico 00791}

\date{\today}

\begin{abstract}
We consider a Kerr black hole acting as a gravitational deflector
within the geometrical optics, and point source approximations.
The Kerr black hole gravitational lens geometry consisting of an
observer and a source located far away and placed at arbitrary
inclinations with respect to the black hole's equatorial plane is
studied in the strong field regime.  For this geometry the null
geodesics equations of our interest can go around the black hole
several times before reaching the observer. Such photon
trajectories are written in terms of the angular positions in the
observer's sky and therefore become ``lens equations".  As a
consequence, we found for any image a simple classification scheme
based in two integers numbers: the number of turning points in the
polar coordinate $\theta$, and the number of windings around the
black hole's rotation axis. As an application, and to make contact
with the literature, we consider a supermassive Kerr black hole at
the Galactic center as a gravitational deflector. In this case, we
show that our proposed computational scheme works successfully by
computing the positions and magnifications of the relativistic
images for different source-observer geometries.  In fact, it is
shown that our general procedure and results for the positions and
magnifications of the images off the black hole's equatorial
plane, reduce and agree with well known cases found in the
literature.
\end{abstract}


\preprint{gr-qc/0308023}

 \maketitle

\section{Introduction}
As is well known, the deflection of light by a gravitating body
was one of the first predictions of Einstein's General Theory of
Relativity to be observationally confirmed. Later on, Einstein
himself predicted what is called today a \textit{microlens}: the
momentarily increase in apparent brightness
of a background star as it passes close to a foreground massive body~\cite%
{einstein}. Both the deflection of light, and the change in
apparent brightness of a radiation source by an external
gravitational field, are collectively known as a
\textit{gravitational lens}.

Nowadays, \textit{gravitational lensing} is a very active area of
research, and it has found applications ranging from the search of
extrasolar planets and compact dark matter to estimate the value
of the cosmological parameters \cite{extraplanets, macho,
parameters}.  In most of these applications it is only necessary
to assume that the gravitational field is weak and that the
deflection angle (due to a spherically symmetric body of mass $M)$
can be approximated by: $\delta \phi \approx 4GM/bc^{2}$, where
$b$ is the impact parameter~\cite{wald}. On the other hand, it is
well known that for a Schwarzschild black hole, the deflection
angle diverges as $b\rightarrow 3\sqrt{3}M$, allowing photons to
orbit the black hole many times before reaching the observer. This
gives rise to an infinite set of images at both sides of the black
hole~\cite{gravitation, bhlens}. Notice that in this region the
gravitational field is no longer weak, and the above approximation
fails.

Recently, a paper by Virbhadra and Ellis have renewed interest in
such images, which they called \textit{relativistic
images}~\cite{ellis}. Later on, Bozza~\cite{bozza, bozza2} and
Eiroa, Romero and Torres~\cite{nordstrom} developed an
approximation method for the case of strong {\it
spherically-symmetric} gravitational fields. In fact, by expanding
the deflection angle near the point of divergence, these authors
were able to find analytic expressions for the positions and
magnifications of the resulting relativistic images.
Interestingly, such images were only characterized by the number
of windings around the black hole~\footnote{Apparently the authors
of~\cite{ellis, bozza, bozza2, nordstrom} were not aware of
Ref.~\cite{bhlens} where, not only the relativistic images for a
Schwarzschild black hole were discussed, but analytic
approximations for their positions and magnifications were also
given.}.

Although several authors have began to study gravitational lenses
with a rotating black hole as a gravitational deflector
(see~\cite{sereno} and references therein), most of their
approaches to this subject still use the weak field approximation,
and focus only in null geodesic motion at the black hole's
equatorial plane. In passing, we shall mention for the interested
reader a nice discussion of a Kerr black hole as a gravitational
lens by Bray~\cite{bray}. Unfortunately, in this work there are
approximations that are valid only for small deviations from the
straight line path, and therefore are not suited for studying
relativistic images. If we want to consider the phenomenology of
the relativistic images in the strong gravitational field of a
Kerr black hole, no approximations can be taken and we need to
work with the full equations of motion for null rays.

There have been numerous articles about the motion of null rays in
the gravitational field of a Kerr black
hole~\cite{cunningham,karas,viergutz1,viergutz2}. Although some of
them have address the gravitational lens problem, most have
concentrated in the observational effects on accretion disks and
sources orbiting at the equatorial plane of the black hole and
have not discussed the phenomenology of the relativistic images.

Recently, Bozza~\cite{bozzaeq} studied ``quasi-equatorial" orbits
of photons around a Kerr black hole and provided analytical
expressions for the positions and magnifications of the
relativistic images. However, these approximations fail when the
observer and the source are located far away from the equatorial
plane. Moreover, as we show in Sec. V, Bozza's procedure begins to
fail when the black hole's angular momentum increases near its
maximum value, even if the observer and the source are close to
the rotating black hole's equatorial plane.

In this paper, we shall discuss the phenomenology of the
relativistic images by using the exact null equations of motion,
with the assumption that the observer and the source are far away
from the black hole. Although this new procedure is far more
complicated than previous works on this subject, it allow us to
calculate, estimate, and discuss for the first time the
observational properties of the relativistic images for arbitrary
source and observer inclinations in a Kerr gravitational lens.

Therefore the purpose of this paper is two-fold. First, to extend
the study of relativistic images for the case when the
gravitational deflector is a rotating Kerr black hole. Secondly,
since the trajectory of a photon will not always will be confined
to a plane, we shall also be concerned with photon trajectories
off the black hole's equatorial plane. As a consequence of this
undertaking, we show below that all relativistic images deflected
by a rotating Kerr black hole are characterized by only two
integers numbers: namely, the number of turning points in the
polar coordinate $\theta$, and the number of windings around the
black hole's rotation axis.

To facilitate reading this paper is divided as follows: in Sec. II
to make this study self-contained a review of the null geodesic's
equations of motion in the Kerr space-time is briefly discussed .
In Sec. III we explain the gravitational lens geometry, and
present a general classification for all images that allow us to
define formally what we mean by ``relativistic images". In Sec.
IV, analytical expressions for the images' magnification are
derived. In Sec. V, we calculate the positions and magnifications
of the relativistic images for several special cases, and compare
our results with those in the literature. Finally, in Sec. VI a
discussion of our results is undertaken. An appendix is also
included to show how the null equations of motion can be solved in
terms of elliptic integrals.

\section{  null geodesics in the kerr space-time}

In this section we briefly review the equations of motion for
light rays in the Kerr space-time. We use the usual Boyer-Lidquist
coordinates which, at infinity, are equivalent to the standard
spherical coordinates. We also discuss about the relevant range of
coordinates and constants of motion for the case of an observer
and a source located far away from the black hole. Details about
the relation between the null geodesic equations, the
gravitational lens geometry, and observable quantities are given
in the Sec. III.

By a convenient choice of the affine parameter, null geodesics in
Kerr space-time can be described by the following first-order
differential system~\cite{carter}:

\begin{eqnarray} \label{ur} \Sigma\, U^r &=& \pm
\sqrt{R(r)}\;,
\\ \nonumber\\
\label{utheta} \Sigma\, U^\theta &=& \pm \sqrt{\Theta(\theta)}\;,
\\ \nonumber\\
\label{uphi} \Sigma\, U^\phi &=&  -\left(a - \frac{\lambda}{\sin^2
\theta}\right) + \frac{a P}{\Delta}\;,
\\\nonumber\\
\label{ut} \Sigma\, U^t &=& -a(a  \sin^2 \theta - \lambda) +
\frac{(r^2+a^2)P}{\Delta}\;,
\end{eqnarray}

where

\begin{eqnarray}
R(r)  &=& r^4  + (a^2 - \lambda^2 - \eta) r^2  \nonumber \\
 &&+ 2\left[(a- \lambda)^2 + \eta \right] r  - a^2 \eta
\;,
\\ \nonumber\\
\label{Theta}
 \Theta(\theta) &=& \eta + a^2 \cos^2\theta -
\lambda^2 \cot^2\theta  \;,
\\ \nonumber\\
P &=& (r^2 + a^2) - \lambda a\;,
\\ \nonumber \\
\Sigma &=& r^2 + a^2 \cos^2\theta \;,
\\ \nonumber\\
\Delta &=& r^2 - 2 r + a^2\;.
\end{eqnarray}

In the above, $U^{\mu }=dx^{\mu }/d\tau $ is the four-velocity
($\tau $ is an affine parameter), $\lambda $ and $\eta $ are the
constants of motion, $a=J/M$ is the black hole's angular momentum
per unit mass, and units are chosen such as $GM/c^{2}=1$.  As is
well known, for a Kerr black hole, the parameter $a$ is restricted
to $0\leq |a|\leq 1$. From Eqs.~(\ref{ur}) - (\ref{ut}) it follows
that the relevant integrals of motion are

\begin{eqnarray}
\label{firstint}
&&\int^r \frac{dr}{\pm \sqrt{R(r)}} = \int^\theta
\frac{d\theta}{\pm \sqrt{\Theta(\theta)}}\;,
\\ \nonumber\\
\label{secondint} \Delta\phi &=& \int^r \frac{a(2r - a
\lambda)}{\pm \Delta \sqrt{R(r)}}\,dr + \int^\theta
\frac{\lambda}{\pm \sin^2\theta \sqrt{\Theta(\theta)}}\,
d\theta\;,
 \nonumber \\ \\
\label{tint} \Delta t &=& \int^r \frac{r^2(r^2 + a^2) + 2a r (a -
\lambda)}{\pm\Delta \left[R(r)\right]^2}\,dr \nonumber \\
&& + \int^\theta \frac{a^2 \cos^2 \theta }{\pm
\sqrt{\Theta(\theta)}}\,d\theta\;.
\end{eqnarray}

The signs of $\sqrt{R(r)}$ and $\sqrt{\Theta (\theta )}$ are those
of $U^{r}$ and $U^{\theta }$ respectively. Thus, the positive sign
is chosen when the lower integration limit is smaller than the
upper limit, and the negative sign otherwise.

For an observer and a source located far away from the black hole,
the relevant radial integrals can be written as follows:
\begin{eqnarray}
\int^{r}\rightarrow -\int_{r_{s}}^{r_{\min }}+\int_{r_{\min
}}^{r_{o}}\approx 2\int_{r_{\min }}^{\infty }\;,
\end{eqnarray}
where $r_{\min }$ is the only turning point in the photon's
trajectory, and it is defined by the largest positive root of
$R(r)=0$. For the angular integrals, however, there could be more
than one turning point. In consequence, the most general
trajectory is described by
\begin{eqnarray}
\int^{\theta }\rightarrow \pm \int_{\theta _{s}}^{\theta _{\min
/\max }}\pm \int_{\theta _{\min /\max }}^{\theta _{\max /\min
}}\cdots \pm \int_{\theta _{\max /\min }}^{\theta _{o}}\;.
\end{eqnarray}
The turning points in $\theta $ are defined by $\Theta (\theta
)=0$ and are given by

\begin{eqnarray}
\label{um}
 \cos^2 \left( \theta_{\min / \max} \right)&=& \frac{1}{2}
\left\{1 - \frac{\lambda^2 + \eta}{a^2} \right.\nonumber \\
&&\left. + \left[\left(1 - \frac{\lambda^2 + \eta}{a^2}\right)^2 +
4\frac{\eta}{a^2} \right]^{1/2} \right\}\;. \nonumber \\
\end{eqnarray}

It is easy to show from Eq.~(\ref{utheta}) that for any pair of
parameters $(\lambda , \eta )$, the motion in $\theta $ is bounded
by $\theta _{\max }$ and $\theta _{\min }$. Thus, the space of
parameters is restricted because $\theta _{\min }\leq \theta
_{o},\theta _{s}\leq \theta _{\max }$. This constrain will be
implemented in Sec. III after discussing the lens geometry.
Finally, the deflection angle $\Delta \phi $ has to be chosen to
satisfy a given source-observer geometry (see Sec. III for
details).

Now we want to know what region of the parameter space ($\lambda
,\eta $) correspond to photons that after reaching $r_{\min }$ can
escape to infinity. Writing Eq.~(\ref{utheta}) as
\begin{eqnarray}
\label{etacarter}
 \eta = U_\theta^2 - a^2 \cos^2\theta + \lambda^2
\cot^2\theta \;,
\end{eqnarray}
we see that $\eta $ could be negative. However, assuming a photon
crosses the equator ($\theta =\pi /2$), implies that $\eta
=U_{\theta }^{2}\geq 0$. Since in this research we consider a
source behind the black hole and we are mostly interested in
images formed by photons that go around the black hole before
reaching the observer, we will only consider the case of positive
$\eta $.

\begin{figure}[h]
\includegraphics[width=8.40cm, height=5.98cm]{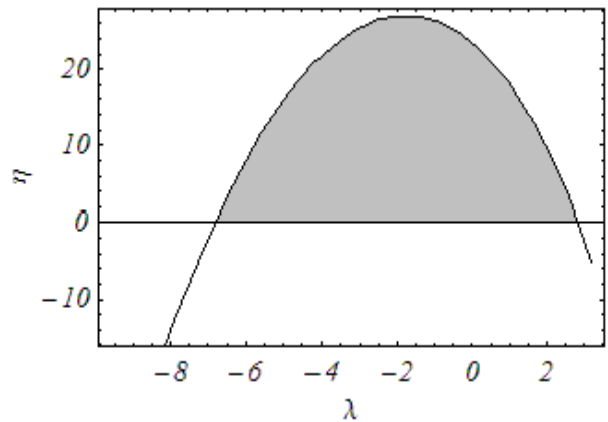}
\caption{\small Photons with parameters ($\lambda,\eta$) inside
the shaded region will fall into the black hole. The parametric
curve is for $a = 0.9$ and is defined by Eqs.~(\ref{lambdamin})
and (\ref{etamin}).} \label{parspace}
\end{figure}

For a photon to be able to return to infinity, we need
$dU^r/{d\tau} > 0$ at $r = r_{\min}$. Setting $U^r = 0 =
dU^r/{d\tau}$ we get:

\begin{eqnarray}
\label{lambdamin} \lambda(\underline{r}) &=& \frac{\underline{r}^2
(\underline{r} - 3) + a^2(\underline{r}+1)}{a(1-\underline{r})}\;,
\\ \nonumber \\
\label{etamin} \eta(\underline{r}) &=& \frac{\underline{r}^3
\left[ 4a^2 - \underline{r} (\underline{r} - 3)^2 \right]}{a^2(1 -
\underline{r})^2}\;,
\end{eqnarray}
where $\underline{r}$ is the lower bound of $r_{\min }$ with
$r_{h}\leq \underline{r}<\infty $, and $r_{h}=1+\sqrt{1-a^{2}}$ is
the Kerr black hole's horizon. This is a parametric curve in the
($\lambda ,\eta $) space and it is shown in Fig.~\ref{parspace}
for the case $a=0.9$. Photons with constants of motion inside the
shaded region do not have a turning point outside the horizon so
they will fall into the Kerr black hole.
This is analogous to the motion of null geodesics around a
Schwarzschild black hole, where any photon with impact parameter
$b\leq 3\sqrt{3}$ can not escape to infinity \cite{ellis,bozza}.
This value of the parameter $b$, correspond to $r_{\min }=3$ which
defines the ``photon sphere". In the parameter space ($\lambda
,\eta $), this correspond to a close region bounded by the line
$\eta =0$ and the curve, $\sqrt{\lambda ^{2}+\eta }=3\sqrt{3}$.
This forbidden region in the parameter space shall be called: {\it
photon region}. The radial integrals of Eqs.~(\ref{firstint}) and
(\ref{secondint}) diverge at the boundary of the photon region
(except at the line $\eta =0$) and take complex value inside of
it. Therefore, knowledge of the mapping of this region will allow
us to ensure that the equations of motion will remain well
behaved.

\section{the kerr gravitational lens geometry}

\begin{figure}[h]
\includegraphics[width=8.64cm,height=7.67cm]{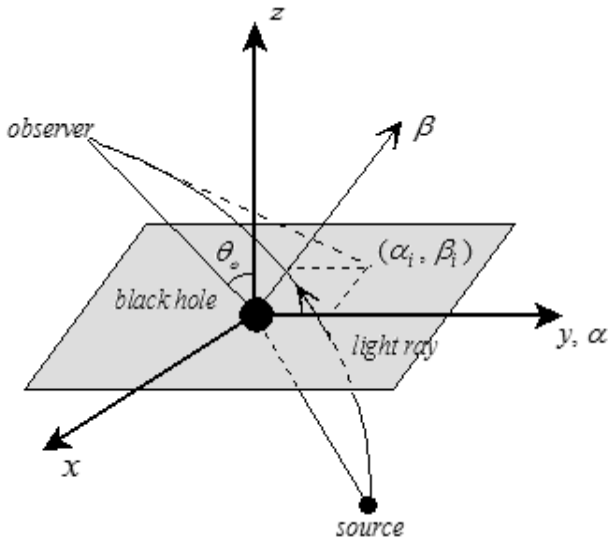}
\caption{\small Geometry of the rotating gravitational lens. An
observer far away from the black hole, can set up a reference
coordinate system $(x,y,z)$ with the black hole at the origin. The
Boyer-Lidquist coordinates coincide with this system only at
infinity. The reference frame is chosen so that, as seen from
infinity, the black hole is rotating around the $z$ axis. In this
system, the line joining the origin with the observer is normal to
the $\alpha$-$\beta$ plane. The tangent vector to an incoming
light ray defines a straight line, which intersects the
$\alpha$-$\beta$ plane at the point ($\alpha_i,\beta_i$).}
\label{lensgeometry}
\end{figure}

Since Kerr space-time is asymptotically flat, an observer far away
from the black hole ($r_{o}\gg 1$) can set up a reference
euclidean coordinate system ($x,y,z$) with the black hole at the
origin (see Fig.~\ref{lensgeometry}). The Boyer-Lidquist
coordinates coincide with this reference frame \textit{only for
large $r$}. The coordinate system is chosen so that, as seen from
infinity, the black hole is rotating around the $z$ axis. For
$a>0$ the rotation will be assumed to be in the counterclockwise
direction as seen from the positive $z$ axis. Without loss of
generality, we choose $\phi _{o}=0$ so that the coordinates of the
observers in the Boyer-Lidquist system are ($r_{o},\theta
_{o},0$). Similarly, for the source we have ($r_{s},\theta
_{s},\phi _{s}$).
In the observer's reference frame, an incoming light ray is
described by a parametric curve $x(r)$, $y(r)$ , $z(r)$, where
$r^{2}=x^{2}+y^{2}+z^{2}$. For large $r$ this is just the usual
radial coordinate in the Boyer-Lidquist system. At the location of
the observer, the tangent vector to the parametric curve is given
by: $\left. (dx/dr)\right\vert _{\,r_{o}}\mathbf{\hat{x}}+\left.
(dy/dr)\right\vert _{\,r_{o}}\mathbf{\hat{y}}+\left.
(dz/dr)\right\vert _{\,r_{o}}\mathbf{\hat{z}}\;.$ This vector
describes a straight line which intersect the $\alpha $-$\beta $
plane shown in Fig.~\ref{lensgeometry} at ($\alpha _{i},\beta
_{i}$). A line joining the origin with the observer is normal to
this plane. We call this plane the \textit{observer's sky}. The
point ($\alpha _{i},\beta _{i}$) in this plane is the point
($-\beta _{i}\cos \theta _{o},\alpha _{i},\beta _{i}\sin \theta
_{o}$) in the ($x,y,z$) system. Changing to spherical coordinates
and using the equations of the straight line, is easy to show that

\begin{eqnarray}
\label{alphai}
 \alpha_i = -r_o^2 \sin\theta_o \left.
\frac{d\phi}{dr} \right|_{r_o} \;,
\\ \nonumber \\
\label{betai}
 \beta_i = r_o^2 \left. \frac{d\theta}{dr}
\right|_{r_o}\;.
\end{eqnarray}

By using Eqs.~(\ref{ur}) - (\ref{uphi}) in (\ref{alphai}) and
(\ref{betai}) and further assuming $r_o \rightarrow \infty$, it is
possible to relate the constants of motion $\lambda$ and $\eta$ to
the position of the images in the observer's sky:

\begin{eqnarray}
\label{lambda}
 \lambda \approx -\alpha_i \sin \theta_o \;,
\\ \nonumber\\
\label{eta} \eta \approx (\alpha_i^2  - a^2)\cos^2 \theta_o +
\beta_i^2 \;.
\end{eqnarray}
These equations can, in turn, be written in terms of the angles
($x_i,y_i$) in the observer's sky by: $\alpha_i \approx r_o \,
x_i$, $\beta_i \approx r_o\, y_i$. We would like to mention that
Eqs.~(\ref{alphai}) - (\ref{eta}) are not equivalent to Eqs.~(7),
(8), (11) and (12) of Ref.~\cite{bray}. However, our equations
exactly coincide with Eqs.~(28a) and (28b) of
Ref.~\cite{cunningham}. We believe that in Ref.~\cite{bray}, there
is a calculation  mistake in the intersection of the tangent to
the light ray and the $\alpha$-$\beta$ plane. The author in Ref.
\cite{bray} also ignore any contribution of the black hole's spin
when expanding $(d\theta /dr)|_{r_{o}}$ and $(d\phi /dr)|_{r_{o}}
$.

 As can be seen later, it is also useful for our purposes to write the
source's angular coordinates ($\theta _{s},\phi _{s}$) in terms of
its position in the observer's sky. From the geometry of
Fig.~\ref{lensgeometry}, we obtain

\begin{eqnarray}
\label{xs}
 x_s \approx \frac{ r_s \sin\theta_s \cos\phi_s}{r_o -
r_s \left(\sin\theta_o \sin\theta_s \cos\phi_s + \cos\theta_o
\cos\theta_s\right)} \;,
\\ \nonumber \\ \nonumber \\
\label{ys}
 y_s \approx \frac{ r_s \left( \sin\theta_o \cos\theta_s
- \cos\theta_o \sin\theta_s \cos\phi_s\right)}{r_o - r_s
\left(\sin\theta_o \sin\theta_s \cos\phi_s + \cos\theta_o
\cos\theta_s\right)}\;,
\end{eqnarray}
where ($x_{s},y_{s}$) are the angles of the source in the
observer's sky. Inverting Eqs.~(\ref{xs}) and (\ref{ys}) for the
polar and azimuthal angles, it is convenient to consider good
source alignments (e.g. small $x_{s}$ and $y_{s}$). The motivation
for this approximation will become clear when we consider the
magnifications of the relativistic images (defined below). If the
observer is not exactly at $\theta _{o}=0$, we can approximate:
$\theta _{s}\approx \pi -\theta _{o}-\delta \theta $ and $\phi
_{s}\approx \pi -\delta \phi $ with $\delta \theta \ll 1$ and
$\delta \phi \ll 1$ (an observer exactly at $\theta _{o}=0$ will
be considered in Sec. V). Expanding Eqs.~(\ref{xs}) and (\ref{ys})
to second order in the perturbations, one finds that

\begin{eqnarray}
\label{thetas} \theta_s &\approx& \pi - \theta_o - \left(1 +
\frac{r_o}{r_s}\right) y_s - \frac{x_s^2
\sin2\theta_o}{4\left(\frac{sin \theta_o}{1 + r_o/r_s} + y_s
\cos\theta_o \right)^2}\;,
\nonumber \\  \\ \nonumber \\
\label{phis} \phi_s &\approx& \pi - \frac{x_s}{\frac{sin
\theta_o}{1 + r_o/r_s} + y_s \cos\theta_o}\;.
\end{eqnarray}

Now we want to know how the restrictions in the parameter space
($\lambda, \eta$) and in the lens geometry are reflected in the
possible values of the image coordinates ($\alpha_i, \beta_i$). We
begin with the photon region discussed in Sec. II. This constraint
correspond to a closed region in the observer's sky. It is useful
to write its boundary as a parametric curve
$\xi(\varphi)$, where $\xi^2 \equiv x^2 + y^2$ and $\tan \varphi \equiv y/x$%
. Inserting Eqs.~(\ref{lambdamin}), (\ref{etamin}), (\ref{lambda}) and (\ref%
{eta}) in the definition of $\varphi$ (using the small angle approximations $%
\alpha \approx r_o \, x$, $\beta \approx r_o\, y$), we can solve for $%
\underline{r}(\varphi)$ and obtain the desired parametric curve.
When doing so, one encounters a sixth order polynomial in
$\underline{r}$. The largest positive root is valid for $-\pi/2
\leq \varphi \leq \pi/2$ and the second largest positive root for
$\pi/2 \leq \varphi \leq 3\pi/2$.

Another constraint comes from the polar movement of the light
rays. In Sec. II we pointed out that the motion in $\theta$ was
restricted between the turning points $\theta_{\min/\max}$ defined
by Eq.~(\ref{um}).
Therefore, points in the parameters space where the inequality $%
\theta_{\min} \leq \theta_s, \theta_o \leq \theta_{\max}$ is not
satisfied must be discarded.

Now we will prove that if $\lambda$ and $\eta$ are given by Eqs.~(\ref%
{lambda}) and (\ref{eta}), then the inequality is always satisfied for $%
\theta_o$. Using the notation: $u_j \equiv \cos^2 \theta_j$, $w_j
\equiv \sin^2 \theta_j$ ($j$ is any subscript), the above
inequality is equivalent to $u_m \geq u_o$ ($m$ stands for
``min/max"). In this notation, the condition for a turning point
in $\theta$, $\Theta(\theta) = 0$, becomes
\begin{eqnarray}
\label{turneq} \eta + \left(a^2-\eta-\lambda^2\right)u_m - a^2
u_m^2 = 0\;.
\end{eqnarray}
Writing $u_{m}=u_{o}-x$ and substituting into Eq.~(\ref{turneq}),
we get

\begin{eqnarray}
x &=& \frac{1}{2a^2}\left\{ a^2 w_o - (\alpha_i^2+\beta_i^2)
\right.
\nonumber \\
 &&  \left. \pm \sqrt{\left[a^2 w_o -
(\alpha_i^2+\beta_i^2)\right]^2 - 4a^2\beta_i^2 w_o } \;
\right\}\;.
\end{eqnarray}
On the other hand, using Eqs.~(\ref{lambda}) and (\ref{eta}) we
have
\begin{eqnarray}
\label{rad}
 \alpha_i^2 + \beta_i^2 = a^2 u_o + \lambda^2 + \eta \;.
\end{eqnarray}
The ``radius" $\lambda ^{2}+\eta $ must be greater or equal than
the boundary of the photon region defined by
Eqs.~(\ref{lambdamin}) and (\ref{etamin}) and the line $\eta =0$.
It is then easy to show that the minimum value of this ``radius"
is reached when $\eta =0$ and $a=1$. The actual minimum value is
thus $(\lambda ^{2}(\underline{r})+\eta (\underline{r}))_{\min
}=4$. Therefore, by Eq.~(\ref{rad}) we have that $\alpha
_{i}^{2}+\beta _{i}^{2}\geq 4$, and since $0\leq a^{2}w_{o}\leq
1$, it follows the inequality $a^{2}w_{o}-(\alpha _{i}^{2}+\beta
_{i}^{2})<0$ and hence, $x\leq 0$. The equality is satisfied only
for $\beta =0$. We then can conclude that $u_{m}\geq u_{o}$.

\begin{figure}[h]
\includegraphics[width=8.64cm,height=6.98cm]{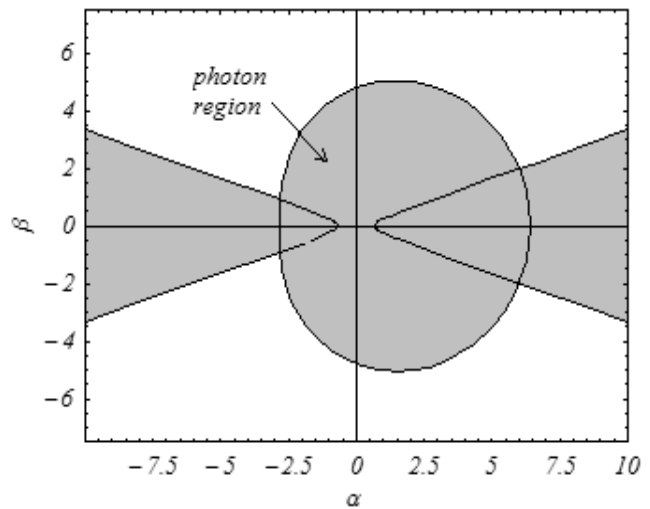}
\caption{\small Forbidden regions in the observer's sky for $u_s
> u_o$. The forbidden regions are shown in shaded gray. In this case we used $a =
1$, $u_o = 1/2$ and $u_s = 1.1 u_o$. For $u_s < u_o$ only the
photon region is present.} \label{freg}
\end{figure}

If $u_s > u_o$, we can ensure that $u_m \geq u_s$ by doing the
same construction in the frame of the source:

\begin{eqnarray}
\label{lambdasource} \lambda^2 = \gamma_i^2 w_s\;, \\
\nonumber\\
\label{etasource} \eta = \Omega_i^2 + u_s(\gamma_i^2 - a^2)\;,
\end{eqnarray}
where $\gamma_i$ and $\Omega_i$ are the coordinates in the
source's sky. By using Eqs.~(\ref{lambda}), (\ref{eta}),
(\ref{lambdasource}) and (\ref{etasource}) it follows that

\begin{eqnarray}
\Omega_i^2 = \beta_i^2 + \alpha_i^2\left(\frac{u_o -
u_s}{w_s}\right)  + a^2(u_s - u_o) \;,\\ \nonumber\\
\gamma_i^2 = \alpha_i^2 \frac{w_o}{w_s}\;.
\end{eqnarray}
Since for physically relevant situations we must have $\Omega_i^2
\geq 0$ (where the equality holds only for $u_m = u_s$), the
excluded points will be those between the curves

\begin{eqnarray}
\label{forbiddencurve}
 \beta_i = \pm \left[ (u_s -
u_o)\left(\frac{\alpha_i^2}{w_s} - a^2 \right)\right]^{1/2} \;,
\end{eqnarray}
where $u_s \geq u_o$ and $\alpha_i^2 \geq a^2 w_s$. The forbidden
regions in the observer's sky are shown in Fig.~\ref{freg} for the
case of $u_s > u_o$. For $u_s < u_o$, only the photon region is
present.

Next, we focus in relating the null equations of motion considered
in Sec. II with the geometry of the gravitational lens system. In
the familiar case of a Schwarzschild black hole lens, the movement
of light rays is restricted to a plane. Therefore, without loss of
generality we can work at the equatorial plane: $\theta =\pi /2$.
Then, by substituting Eq.~(\ref{firstint}) in (\ref{secondint})
and setting $a=0$, we obtain the familiar Schwarzschild deflection
angle:
\begin{eqnarray}
\label{schang}
\Delta\phi = 2 \int_{r_{min}}^\infty \frac{ \lambda
\, dr}{\sqrt{r^4 + \lambda^2 \,r (2 - r)}} \;,
\end{eqnarray}
where $\eta =0$ by the fact that we are working at the equatorial
plane [see Eq.~(\ref{etacarter})]. This deflection angle can be
written in terms $x_{i}$ using Eq.~(\ref{lambda}). Then, by use of
the familiar ``lens equation" we can solve for the position of the
``virtual" images~\cite{ellis}. However, another approach is to
give $\Delta\phi $ in terms of the geometry of the observer-source
pair and solve Eq.~(\ref{schang}) for $x_{i}$. In this case we
have

\begin{eqnarray}
\label{dphi}
 \Delta\phi = \left\{
\begin{array}{ll}
    -\phi_s - 2\pi n &, \;{x_i > 0} \\
    2\pi( n +1) - \phi_s  &,\; {x_i < 0} \\
\end{array}
\right.
\end{eqnarray}
where $n=0,1,2\ldots $ is the number of windings around the $z$
axis. The solutions for $n=0$ are the familiar weak field images,
and for $n\geq 1$ we have the relativistic images studied in
\cite{ellis, bozza, bozza2}. This last method is more useful when
we consider trajectories outside the equatorial plane. In this
case, Eqs.~(\ref{firstint}) and (\ref{secondint}) with
(\ref{dphi}) become our ``lens equations". Nevertheless, for
trajectories outside the equatorial plane, an additional parameter
appears: the number of turning points in the polar coordinate
$\theta $ ($m$). For the case of the Schwarzschild black hole it
is easy to prove that $m$ is related to $n$ as $m=2n+1$ by the
fact that photons travel only in a plane.

What about the case of a Kerr black hole? In this situation, since
the movement of light rays is not necessarily restricted to a
plane, we must consider $n$ and $m$ as independent parameters.
Additionally, in trying to write $\Delta \phi$ as in
Eq.~(\ref{dphi}) one is faced with a possible complication: unlike
in the Schwarzschild space-time, the Kerr geometry admits turning
points in $\phi$ which could complicate the analysis. By using
Eq.~(\ref{uphi}) with $U^{\phi }=0$ is easy to show that the
possible turning points occur at

\begin{eqnarray}
r_{\pm}(\theta) &=& \frac{1}{\lambda} \left\{ (\lambda -
a\sin^2\theta) \right. \nonumber \\
&& \left.\pm \sqrt{(\lambda - a\sin^2\theta)^2 - a^2 \lambda^2
\cos^2\theta } \right\}\;.
\end{eqnarray}

Both solutions are surfaces of revolution around the rotation axis
of the black hole. It is easy to prove that for a given $\lambda
$, only one of the surfaces lies outside the horizon. Since in our
case we are considering photons that come from infinity, reach a
single turning point in $r$ and return to infinity, they can cross
such a surface at most in two points. In consequence, there can be
at most two turning points in $\phi $. However, by using
Eqs.~(\ref{uphi}) and (\ref{ut}) for an observer and a source at
large $r$, it is easy to prove that: $r_{s}^{2}\sin ^{2}\theta
_{s}(d\phi /dt)_{s}\approx \lambda \approx r_{o}^{2}\sin
^{2}\theta _{o}(d\phi /dt)_{o}$ (conservation of angular
momentum). Therefore, the asymptotic sign of $d\phi /dt$ must be
unchanged by the gravitational interaction. This rule out a single
turning point in $\phi $ which would obviously change that sign.
We conclude that the number of turning point in $\phi $ must be
zero or two. In that case, the most general expression for $\Delta
\phi $ is still given by Eq.~(\ref{dphi}) but with the following
modification: the first case is to be use when the right hand side
of Eq.~(\ref{secondint}) is negative, and the second case
otherwise.

By using Eqs.~(\ref{lambda}), (\ref{eta}), (\ref{thetas}) and
(\ref{phis}), the ``lens equations" (\ref{firstint}) and
(\ref{secondint}) can be expressed as

\begin{eqnarray}
\label{lens1} H(x_i,y_i) - G(x_i,y_i,x_s,y_s,m) = 0\;,
\\ \nonumber \\
\label{lens2}\Delta\phi(x_s,y_s,n) - L(x_i,y_i) -
J(x_i,y_i,x_s,y_s,m)=0\;,
\end{eqnarray}
where $H$ and $L$ ($G$ and $J$) are the radial (angular) integrals
of Eqs.~(\ref{firstint}) and (\ref{secondint}) respectively, and
$n = 0,1,2,\ldots$ and $m = 0,1,2,\ldots$ are the number of
windings around the $z$ axis and the number of turning points in
the polar coordinate $\theta$ respectively.

Although the integers $n$ and $m$ should be considered
independent, as we will see in Sec. IV, the magnification does not
depend directly on $n$. Therefore, in this paper we consider $m$
to be the fundamental parameter. This interpretation is confirmed
by our numerical results where we find that the magnification
always decreases as we increase $m$, and for a given $m$, it can
ever increase for images with larger $n$ (see Sec. V).

In the familiar Schwarzschild gravitational lens, we always have
two images which are formed by light rays which suffer small
deviations in their trajectories. They are called \textit{weak
field images}. Additionally, we have an infinite set of faint
images at both sides of the black hole~\cite{ellis}. These are
\textit{relativistic images}, and are the result of photons that
orbit the black hole several times before reaching the observer.
In Kerr space-time, where photons' trajectories are no longer
confined to a plane, the concept of ``orbiting the black hole
several times" can be subtle. For this reason, we classify images
as follows: images with $m=0,1$ are called \textit{direct images}
(DI hereafter), and images with $m\geq 2$ are called
\textit{relativistic images} of order $m$ (RI hereafter).

\begin{figure}[h]
\includegraphics[width=8.64cm,height=9.00cm]{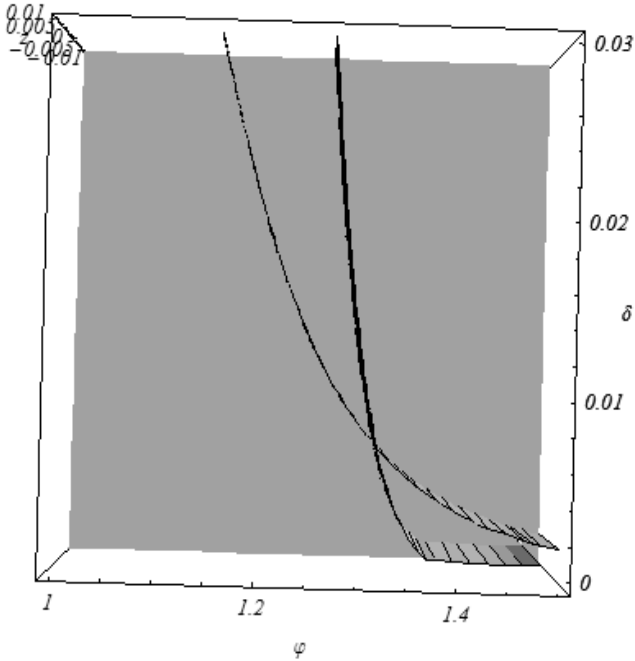}
\caption{\small Here we plot the surfaces $z_1(\delta,\varphi) =
H(\delta,\varphi) - G(\delta,\varphi,x_s,y_s,m)$ and
$z_2(\delta,\varphi) = \Delta\phi(x_s,y_s,n) -L(\delta,\varphi)-
J(\delta,\varphi,x_s,y_s,m)$ for a source at $x_s = y_s =
1/\sqrt{2}$ arcmin,  an observer at $\theta_o = \pi/4$ and for
$n=3$, $m=4$, $a=0.9999$. We have chosen $r_o = r_s = 8.5$ kpc as
in the numerical examples of section V. The intersection of the
surfaces with the $z = 0$ plane (gray) form various curves in that
plane. The intersection of two  such curves marks the position of
a solution to the ``lens equations". The range of $z$ has been
restricted to $-0.01 \leq z \leq 0.01$ in order to clarify the
intersection of the curves. } \label{approxmethod}
\end{figure}

As we show in the Appendix, Eqs.~(\ref{lens1}) and (\ref{lens2})
can be written in terms of elliptic integrals, and are highly
non-linear in all arguments with the exception of $n$ and $m$.
Therefore, their solution require numerical and graphical methods.
To solve the lens equations, we use the standard routine
``FindRoot" build in MATHEMATICA~\footnote{Information about
MATHEMATICA can be found at {\it www.wolfram.com}.}. To give an
initial approximation and to ensure that a solution exist at all,
we use a variety of graphical methods. Since, as we will see in
Sec. V, RIs appear very near the boundary of the photon region, we
found that is useful to write: $x_{i}=(\xi (\varphi )+\delta )\cos
\varphi $ and $y_{i}=(\xi (\varphi )+\delta )\sin \varphi $, where
$0\leq \varphi \leq 2\pi $, $\delta \geq 0$ and $\xi (\varphi )$
is the boundary of the photon region as discussed earlier in this
section. We then express all functions in terms of $\delta $ and
$\varphi $ to avoid entering the photon region. We find that RIs
form at very small $\delta $. To find the images we plot the
surfaces $z_{1}(\delta ,\varphi )=H(\delta ,\varphi )-G(\delta
,\varphi ,x_{s},y_{s},m) $ and $z_{2}(\delta ,\varphi )=\Delta
\phi (x_{s},y_{s},n)-L(\delta ,\varphi )-J(\delta ,\varphi
,x_{s},y_{s},m)$ for a given source position and image numbers
$(n,m)$. The intersection of both surfaces with the plane $z=0$
form various curves in that plane. If the curves formed by the two
surfaces intersect each other, there is a solution. Visual
inspection of these curves allow us to give an initial
approximation for the numerical routine. As an example, we plot
part of both surfaces in Fig.~\ref{approxmethod} for $n=3$ and
$m=4$.
 We usually do this for the two lowest values of $m$ where a solution can be
found. The possible values of $n$ are bounded from above by the requirement $%
|\Delta \phi |\leq \max |L(\delta ,\varphi )+J(\delta ,\varphi
,x_{s},y_{s},m)|$. In consequence, for a given $m$ we usually find
images for just a few values of $n$. This procedure is very
tedious and time consuming, and for these reasons we are unable to
give a complete phenomenological description of the behavior of
the images for any given observer-source geometry. Instead, we
shall present some numerical examples of the kind of behavior that
can be expected by having a Kerr black hole as a gravitational
deflector and null geodesic motion off the equatorial plane.

As pointed out before, several authors have already obtained
analytic approximations to the radial integrals in the ``strong
field limit" for Schwarzschild, Reissner-Norsdstr\o n and Kerr
black holes~\cite{bozza, bozza2, bozzaeq, nordstrom}. Their
approximations are valid for relativistic photons, with $r_{\min
}$ close to $\underline{r}$ and for small deviations from the
equatorial plane in the case of the Kerr black hole. However,
since in this article we are interested in orbits that can deviate
significantly from the equatorial plane we can no longer use such
approximation schemes. Moreover, their whole approach fail for the
angular integrals because, in general, there is no relation
between $r_{\min }$ and the turning points in $\theta$. In Sec. V,
to make contact with the relevant literature, we consider
trajectories close to the equatorial plane to be compared with
Bozza~\cite{bozzaeq} procedures.

\section{The Magnification in a Kerr Gravitational Lens}

The magnification of an image is defined as the ratio of the
observed flux to the flux of the unlensed source. By Liouville's
theorem, the surface brightness is unchanged by the gravitational
light deflection \cite{glens}. Therefore, the magnification is
defined as the ratio of the solid angle subtended by the image to
the solid angle of the unlensed source:

\begin{eqnarray}
\mu \equiv \frac{d\omega_i}{d\omega_s} = \frac{1}{|J|}\;,
\end{eqnarray}
where $J$ is the Jacobian of the transformation ($x_s,y_s$)
$\rightarrow$ ($x_i,y_i$). Writing $x_s = x_s(x_i,y_i)$ and $y_s =
y_s(x_i,y_i)$ we can find expressions for $\partial x_s/\partial
x_i$, $\partial x_s/\partial y_i$, $\partial y_s/\partial x_i$ and
$\partial y_s/\partial y_i$ by differentiating the lens equations
(\ref{lens1}) and (\ref{lens2}) with respect to $x_i$ and $y_i$.
After some algebraic manipulations, we find that

\begin{eqnarray}
\label{mag} \mu = \left| \frac{\alpha_1 \alpha_4 - \alpha_2
\alpha_3}{\beta_1 \beta_4 - \beta_2 \beta_3} \right| \;,
\end{eqnarray}
where

\begin{eqnarray}
\alpha_1 = \frac{\partial G}{\partial x_s}\;,  \;\;\;\; \alpha_2 =
\frac{\partial G}{\partial y_s}\;,
\\ \nonumber \\
\alpha_3 = -\frac{\partial \phi_s}{\partial x_s} - \frac{\partial
J}{\partial x_s}\;,  \;\;\;\; \alpha_4 = -\frac{\partial
\phi_s}{\partial y_s} - \frac{\partial J}{\partial y_s}\;,
\\ \nonumber \\
\beta_1 = \frac{\partial H}{\partial x_i} - \frac{\partial
G}{\partial x_i}\;,\;\;\;\; \beta_2 = \frac{\partial H}{\partial
y_i} - \frac{\partial G}{\partial y_i}\;,
\\ \nonumber \\
\beta_3 = \frac{\partial L}{\partial x_i} + \frac{\partial
J}{\partial x_i}\;, \;\;\;\; \beta_4 = \frac{\partial L}{\partial
y_i} + \frac{\partial J}{\partial y_i}\;.
\end{eqnarray}

The above derivatives are very cumbersome and we will not expand
it, neither show explicitly them here. In fact, instead we shall
use the equivalent numerical derivatives in all forthcoming
calculations. Note that the magnification do not depend on $n$
directly, since we have to take the derivative of $\Delta \phi $.
Because of the complexity of these expressions, we are unable to
give a complete description of the caustic structure of the Kerr
space-time. Thus, we shall limit ourselves to compute the
magnification of the images that we found. For a description of
the caustic structure of a Kerr space-time, see \cite{blandford,
bozzaeq}.

\section{Relativistic Images by a Kerr Black Hole Gravitational Lens.}

In this section we present numerical calculations for the
positions and magnifications of the RIs for different
source-observer geometries. Our purpose is to provide a physical
insight of the phenomenology that can be expected from a rotating
black hole behaving as a gravitational deflector.  Also, our
general procedures and numerical solutions can provide a set of
``test-bed" calculations for more sophisticated gravitational lens
models to be developed in the near future. To be able to compare
our results with recent published articles~\cite{ellis, bozza,
bozza2, bozzaeq} we shall consider a gravitational lens composed
of a rotating black hole at the Galactic center with a mass of
$M=2.8\times 10^{6}M_{\odot }$, and located at a distance of
$r_{o}=8.5$ kpc~\cite{nature}. We shall also take $r_{s}=r_{o}$.

The section is subdivided as follows: we first consider two simple
cases involving a Schwarzschild black hole ($a=0$) and an observer
located at the pole ($\theta _{o}=0$). Next, as a consistency
check, we consider an observer at the equator ($\theta _{o}=\pi
/2$) and compare our calculations with those of
Bozza~\cite{bozzaeq}. Finally, we work out a more general case of
an observer at say $\theta _{o}=\pi /4$. The RIs are classified as
follows: we use a plus (+) sign for images with sign $y_{i}=$ sign
$y_{s}$ (e.g. in the same ``side" of the source) and a negative
($-$) sign otherwise. We do all calculations for the two lowest
values of $m$ where a solution exist. We find that, in general,
images with larger $m$ are more demagnified. The following
notation will also be used: $\xi \equiv \sqrt{x^{2}+y^{2}}$,
$\varphi \equiv \arctan (y/x)$, where $(x,y)$ are the angular
positions in the observer's sky.

In all geometric configurations considered in this section, we
found that for $m=1$ we recovered the usual weak field images
without any noticeable effect from the black hole's spin.
Moreover, we could not find any image with $m=0$ for this kind of
geometry.

\subsection{A Schwarzschild Black Hole as a Gravitational Deflector}

In the Appendix we show that for $a=0$, the angular integrals can
be solved in close form. As expected, we find that the RIs are
always found in the line joining the source position $(x_s, y_s)$
and origin of the observer's sky regardless of the inclination of
the observer $\theta_o$. The separation $\xi_i$ of the two
outermost images is always the same regardless of the source
position (or $\theta_o$). They are: $\xi_+ = \xi_- \approx 16.952
\; \mu$arcsec ($m=3$) and $\xi_+ = \xi_- \approx 16.931 \;
\mu$arcsec ($m=5$) for the two lowest order images. These results
are close to those found in~\cite{ellis}: $\xi_+ = \xi_- \approx
16.898 \; \mu$arcsec and $\xi_+ = \xi_- \approx 16.877 \;
\mu$arcsec respectively. The magnifications calculated by
Eq.~(\ref{mag}) are also close to those  in the literature.
However, we encountered problems of numerical noise when using the
exact expression for $J(x_i,y_i,x_s,y_s,m)$. For that reason, to
check that we obtain the right magnifications, we used the
expression for $a \neq 0$ [Eq.~(\ref{Jelliptic})] and set $a
\approx 10^{-11}$. The magnifications obtained this way for the
outermost image as a function of the source separation agree with
the literature~\cite{ellis}. For instance, we obtained $\mu_+ =
\mu_- \approx 3.5 \times 10^{-18}$ and $\mu_+ = \mu_- \approx 3.6
\times 10^{-14}$ for $\xi_s = 1$ arcsec and $\xi_s = 100\;
\mu$arcsec respectively. For comparision, in Ref.~\cite{ellis} we
read: $\mu_+ = \mu_- \approx 3.5 \times 10^{-18}$ and $\mu_+ =
\mu_- \approx 3.5 \times 10^{-14}$ respectively.

\subsection{An Observer at the Kerr Black Hole's Pole ($\protect\theta_{o}=0 $) }

Although is very unlikely that an observer will be exactly at
$\theta_o = 0$, this is the simplest case that can be solved
quasi-analytically. Using Eq.~(\ref{lambda}) for $\theta_o=0$ we
find that $\lambda \approx 0$. The minimum value of $\eta$,
corresponding to the photon sphere, is found  from
Eqs.~(\ref{lambdamin}) and (\ref{etamin}) by setting $\lambda =
0$. We obtain

\begin{eqnarray}
\label{etaminpolar}
 \eta_{\min}(a) =  \frac{\underline{r}(a)^2
(3\underline{r}(a)^2 + a^2)}{\underline{r}(a)^2 - a^2}\;,
\end{eqnarray}
where
\begin{eqnarray}
\underline{r}(a) = \frac{1}{3} (3 - a^2)\left[1 + 2 \cos(\psi/3)
\right]\;, \\ \nonumber \\
\tan \psi = \frac{3\sqrt{3} a \sqrt{108 - 135 a^2 + 36 a^4 - 4
a^6}}{54-81a^2 + 18a^4 - 2a^6}\;.
\end{eqnarray}
Eq.~(\ref{etaminpolar}) is a decreasing function of $a$ satisfying
$14.33 \leq \eta_{\min} \leq 27$. Since we can relate the position
of the images to $\eta$ by Eq.~(\ref{eta}): $\xi_i \approx
(1/r_o)\sqrt{\eta + a^2}$, Eq.~(\ref{etaminpolar}) sets a lower
bound to the separation of the images $\xi_i$.

\begin{figure}[h]
\includegraphics[width=8.64cm,height=6.06cm]{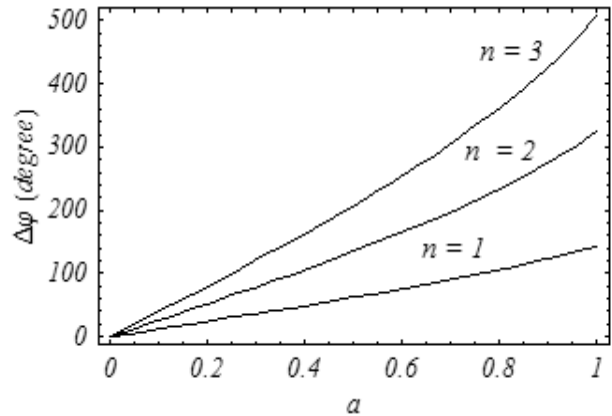}
\caption{\small Deflection in the image inclination ($\Delta
\varphi \equiv \varphi_i - \varphi_s$) as a function of the
normalized black hole angular momentum $a$. Here we consider an
observer located at $\theta_o = 0$ and a source located at $\xi_s
= 1\;\mu$arcsec. The parameter $n$ is the number of loops around
the black hole. These plots are for $p= -1$, but photons with $p =
1$ have almost the same curves (indistinguishable within the plot
resolution).} \label{deltavarphi}
\end{figure}

\begin{figure}[h]
\includegraphics[width=8.51cm,height=9.05cm]{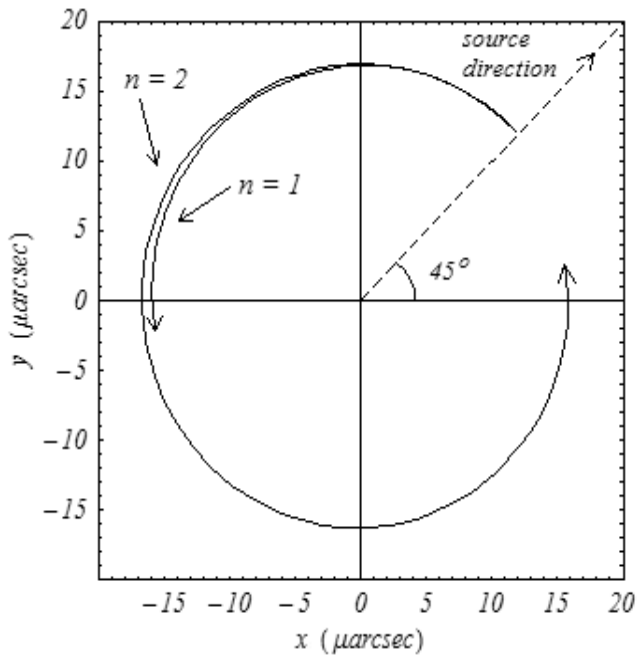}
\caption{\small Positions of the two lowest order primary RIs ($n
= 1,2$ and $p = -1$) as a function of the normalized angular
momentum of the black hole $a$. The arrows in the curves represent
the direction of the movement as we increase $a$ from 0 to 1. The
source is located at $\xi_s = 1\;\mu$arcsec and $\varphi_s =
\pi/4$.} \label{polarmovement}
\end{figure}

The null equations of motion simplify considerable for $\lambda
=0$. First we note that from Eq.~(6), there are no turning points
in $\theta$ since $\eta >0$. Therefore, the angular integration
limits can be shown to be
\begin{eqnarray} p \int_0^{\pi -
\theta_s} + 2(2n+1)\int_0^{\pi/2}\;,
\end{eqnarray}
where $p = \pm 1$ is the parity of the image ( $-$ for images that
are in the same side of the source when $a= 0$ and $+$ for images
in the opposite side). Here, $n$ is the number of loops around the
black hole. To relate $\theta_s$ to the unperturbed position of
the source in the observer's sky we can no longer use the
approximate expressions (\ref{thetas}) and (\ref{phis}). Instead,
it can be shown that

\begin{eqnarray}
\theta_s \approx \pi - \xi_s\left(1 + \frac{r_o}{r_s}\right)\;,
\end{eqnarray}
where we are assuming that $\xi_s \left(1 + r_o/r_s\right) \ll 1$.
Now, using Eqs.~(\ref{helliptic}), (\ref{lelliptic}) and
(\ref{int2}) given in the Appendix, the null equations of motion
(\ref{firstint}) and (\ref{secondint}) become:

\begin{eqnarray}
\label{firstintpolar}
&&\frac{4}{\sqrt{(r_a - r_c)(r_b - r_d)}} \,F(\psi,k) \nonumber \\
&&\;\; \;\;\;\;\;\;\; -\frac{1}{r_o \xi_i} \left\{p \,F
\left[\xi_s\left(1 + \frac{r_o}{r_s}\right), \frac{a^2}{r_o^2
\xi_i^2}\right] \right.
\nonumber \\
&& \;\; \;\;\;\;\;\;\; \;\;\;\;\; \left. + 2 (2n+1)\, K
\left(\frac{a^2}{r_o^2 \xi_i^2}\right) \right\} = 0\;,
\end{eqnarray}
\begin{eqnarray}
\label{secondintpolar}
 \Delta \varphi =\frac{4 a}{\sqrt{(r_a -
r_c)(r_b - r_d)}} &&\sum_{j = 1}^{2} \Gamma_j \left[ (1 - \beta_j)
\Pi(\psi,\alpha_j^2,k)\right. \nonumber \\
&& \left. + \beta_j F(\psi,k)\right]\;,
\end{eqnarray}
where

\begin{eqnarray}
\psi = \arcsin \sqrt{\frac{r_b- r_d}{r_a - r_d}} \;,
\\ \nonumber \\
k^2 = \frac{(r_b - r_c)(r_a - r_d)}{(r_a - r_c)(r_b - r_d)}\;,
\\ \nonumber \\
\alpha_i^2 = \frac{(r_j  - r_b)(r_a - r_d)}{(r_j - r_a)(r_b -
r_d)}\;,
\\ \nonumber \\
 \Gamma_j = \frac{\sqrt{1 - a^2} +
(-1)^{j+1}}{\sqrt{1- a^2}\left((-1)^j \sqrt{1 - a^2} + r_a - 1
\right)}\;,
\\ \nonumber \\
\beta_j = \frac{r_j - r_a}{r_j - r_b}\;,
\\ \nonumber \\
r_j = 1 + (-1)^{j+1}\sqrt{1 -a^2}\;,
\end{eqnarray}
and $r_a = r_{\min}$, $r_b$, $r_c$, $r_d$ ($r_a > r_b > r_c >
r_d$) are the roots of $R(r) = 0$, and Eq.~(\ref{firstintpolar})
is valid only for $a \neq 1$ (for $a = 1$ the right hand side of
Eq.~(\ref{secondintpolar}) has to be substituted by
Eq.~(\ref{lelliptica0}) of the Appendix). The procedure to
calculate the positions of the RIs is the following: for a given
source separation $\xi_s$, we use Eq.~(\ref{firstintpolar}) to
calculate the angular separation of the RI ($\xi_i$) and then
insert this value in Eq.~(\ref{secondintpolar}) to obtain the
offset from the source inclination $\Delta\varphi  \equiv
\varphi_i  - \varphi_s$. In calculating $\xi_i$ one has to use
Eq.~(\ref{etaminpolar}) to determine its minimum value and avoid
numerical problems.  Note that for $a = 0$,
Eq.~(\ref{firstintpolar}) gives $\varphi_i = \varphi_s$ as
expected. In Fig.~\ref{deltavarphi} we plot $\Delta\varphi$ for
the three outermost images ($n = 1,2,3$) as a function of $a$ for
a source located at $\xi_s = 1\;\mu$arcsec.
The sign of $\Delta\varphi$ is that of $a$ (which is what one
intuitively expects). For a given $n$, both images with opposite
parity have almost the same $\Delta\varphi$. However, the greater
the number of loops $n$, the greater the deflection
$\Delta\varphi$ for a given $a$.  The physical picture emerging is
a very simple one: the angular momentum of the black hole just
adds a ``twist" in the direction of the rotation to the usual
Schwarzschild trajectory. To give an intuition of the full
movement of the images, including their separation $\xi_i$, in
Fig.~\ref{polarmovement} we have plotted the position of the two
lowest order images ($n = 1,2$) as a function of the spin
parameter $a$.

\begin{figure}[h]
\includegraphics[width=8.71cm,height=5.58cm]{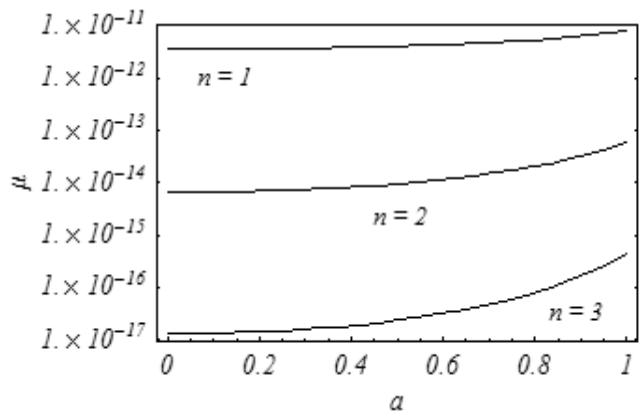}
\caption{Magnification for the three lowest order RIs ($n =
1,2,3$) for an observer at $\theta_o = 0$ and a source at $\xi_s =
1 \mu$arcsec. Here we consider the case of $p = -1$, but the
magnifications for $p = 1$ are almost the same (indistinguishable
in this plot).} \label{magpolarplot}
\end{figure}

What about the magnifications? In this simple case where we have
circular symmetry as seen by the observer the magnification is
given by

\begin{eqnarray}
\label{magpolar}
 \mu &=& \left| \frac{\xi_i}{\xi_s}
\frac{d\xi_i}{d\xi_s} \right|
\nonumber \\
&=& \frac{\xi_i}{\xi_s} \left| \left(\frac{\partial
g}{\partial\xi_s}\right) \left(\frac{\partial g}{\partial
\xi_i}\right)^{-1} \right|\;,
\end{eqnarray}
where $g = g(\xi_i,\xi_s)$ is the left hand side of
Eq.~(\ref{firstintpolar}). This expression for the magnification
can be verified by calculating the Jacobian of the transformation
$(\xi_s,\varphi_s) \rightarrow (\xi_i,\varphi_i)$. We find that,
as in the Schwarzschild case, both images with the same winding
number $n$ have approximately the same magnification (to at least
three significant figures). In Fig.~\ref{magpolarplot} we plot the
magnifications for the three lowest order images ($n = 1, 2, 3$)
and with a source located at $\xi_s = 1\;\mu$arcsec.
The net effect of the angular momentum in this case is to enhance
the brightness of the images. For the Schwarzschild case ($a =
0$), the magnifications obtained with Eq.~(\ref{magpolar}) agree
perfectly with the results of Ref.~\cite{ellis}.

\subsection{ An Observer at Kerr Black Hole's Equator ($\protect\theta _o=\pi /2$) }

In this subsection we consider an observer located exactly at
$\theta_o = \pi/2$. Our purpose is not to give a complete account
of the phenomenology of the RIs since, from an astronomical
perspective, it is very unlikely that an observer will be exactly
at the equator (also setting $\theta_o = \pi/2$ does not
significantly simplify our analysis). Rather than that, this
special case will serve as a consistency check by allowing us to
compare some of our numerical results with those of
 Bozza~\cite{bozzaeq}.

 In that article, the author considered
``quasi-equatorial" orbits in Kerr spacetime. One of the main
approximations employed  was that the horizontal position of the
RIs ($x_i$ in our notation) was calculated independently using the
familiar lens equation in the equatorial plane. This, of course,
assumed that the motion in $\phi$ was unaffected by the motion in
$\theta$. Therefore, we are forced to consider very small source
declinations $\varphi_s \ll 1$. To this end (and for simplicity)
we fix the source at $x_s = 1$ arcsec and $y_s = 1 \;\mu$arcsec
($\varphi_s \approx 10^{-6}$). Then, we calculate the position and
magnification of the lowest order RI in the same side of the
source as a function of the spin parameter $a$. We fixed $n = 1$
to be consistent with the reference, where the RIs are classified
only with the winding number $n$. We also use $a < 0$ for latter
convenience. The results using the approximate equations (66),
(77) and (90) of Ref.~\cite{bozzaeq} and the equations of this
paper are shown in Table I \footnote{Note that in
Ref.~\cite{bozzaeq} the author use units where $2GM/c^2 = 1$ and,
in their notation, $a > 0$ is equivalent to $a < 0$ in our
notation.}.
\begin{table*}
  \caption{Here we fix the source at $\xi_s = 1$ arcsec, $\varphi_s = \pi/4$ and calculate the position
  and magnification of the lowest order RI in the same quadrant as the source as a function of the black
   hole normalized angular momentum $a$.
  The numerical results using the approximate expressions of
  Ref.~\cite{bozzaeq} are shown in columns 2 - 4. Our results are shown in columns 5 - 8. We also show the number of
  turning points for each case ($m$). All angular positions are given in microarcseconds and all angles in radians.}
\begin{tabular}{c | c c c |c c c c}
\hline\hline
    & \multicolumn{3}{c}{using the formulas of Bozza} \vline &\multicolumn{4}{c}{using the formulas of this paper} \\\hline
   $a$ & \;\;\;\;\; $\xi_i$ \;\;\;\;\; &\;\;\;\;\; $\varphi_i/\pi$ \;\;\;\;\;& \;\;\;\;\; $\mu$ \;\;\;\;\; &
   \;\;\;\;\; $\xi_i$ \;\;\;\;\; &\;\;\;\;\; $\varphi_i/\pi$ \;\;\;\;\;& \;\;\;\;\; $\mu$ \;\;\;\;\;  &  $m$  \\
  \hline
  $-10^{-6}$ & 16.952 & $2.615\times10^{-7}$ & $2.9\times10^{-18}$ & 16.952 & $2.618\times10^{-7}$ & $2.9\times10^{-18}$&3 \\
  $-0.1$ & 16.295 & $1.442\times10^{-11}$ & $1.8\times10^{-22}$& 16.296 & $1.444\times10^{-11}$ & $1.8\times10^{-22}$&3 \\
  $-0.2$ & 15.619 & $7.161\times10^{-12}$ & $1.0\times10^{-22}$& 15.619 & $7.174\times10^{-12}$ & $1.1\times10^{-22}$&3  \\
  $-0.3$ & 14.917 & $4.832\times10^{-12}$ & $8.1\times10^{-23}$& 14.918 & $4.844\times10^{-12}$ & $8.5\times10^{-23}$&3  \\
  $-0.4$ & 14.187 & $3.758\times10^{-12}$ & $7.4\times10^{-23}$& 14.188 & $3.771\times10^{-12}$ & $7.6\times10^{-23}$&3  \\
  $-0.5$ & 13.422 & $3.223\times10^{-12}$ & $7.5\times10^{-23}$& 13.423 & $3.240\times10^{-12}$ & $8.1\times10^{-23}$&3  \\
  $-0.6$ & 12.612 & $3.024\times10^{-12}$ & $8.5\times10^{-23}$& 12.614 & $3.050\times10^{-12}$ & $1.8\times10^{-22}$&3  \\
  $-0.7$ & 11.743 & $3.155\times10^{-12}$ & $1.1\times10^{-22}$& 11.747 & $3.207\times10^{-12}$ & $1.2\times10^{-22}$&2  \\
  $-0.8$ & 10.790 & $3.881\times10^{-12} $ & $1.6\times10^{-22}$& 10.802 & $4.051\times10^{-12}$ & $1.7\times10^{-22}$&2  \\
  $-0.9$ & 9.695  & $6.672\times10^{-12}$ & $3.3\times10^{-22}$& 9.741 & $8.428\times10^{-12}$ & $5.0\times10^{-22}$&2  \\
  $-0.95$ & 9.018 & $9.763\times10^{-12}$ & $4.6\times10^{-22}$& 9.147 & $3.672\times10^{-11}$ & $2.8\times10^{-21}$&2  \\
  $-0.99$ & 8.063 & $4.058\times10^{-12}$ & $9.7\times10^{-23}$& 9.042 & $0.1167$ & $7.0\times10^{-22}$&2  \\
\hline\hline
\end{tabular}
\end{table*}

\begin{figure}[h]
\includegraphics[width=8.93cm,height=5.08cm]{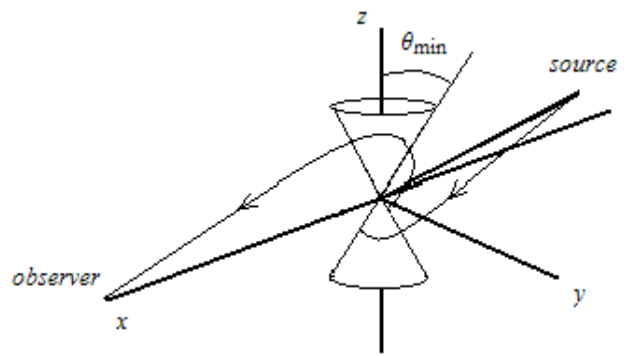}
\caption{A typical light ray trajectory with $m = 2$ and $n = 1$
for an observer at the equator ($\theta_o = \pi/2$). The cones
represent the turning point angles $\theta_{\min /\max}$. The
number of turning points in a trajectory is the number of times
the photon touches these cones (in this case, two). The drawing is
not to scale.} \label{trajectory}
\end{figure}

The general behavior of the position of this image for $a < 0$ is
the following: as we increase $|a|$, it moves toward the $x$ axis
and gets very close to the forbidden region defined by Eq.
(\ref{forbiddencurve}). Then for $|a| \gtrsim 0.7$ it starts to
increase $\varphi_i$ until a maximum value of $\varphi_i \approx
0.1 \pi$ is reached (when $|a| \rightarrow 1$). This increase in
$\varphi_i$ coincides with a change in the order of the image
around $a  \sim 0.6 - 0.7$: from $m = 3$ to $m = 2$. This
illustrates the importance of considering $m$ as an independent
parameter. Also, it gives a better physical intuition of the
qualitative form of the photon's trajectory (see
Fig.~\ref{trajectory}).

The calculations done with the approximate expressions of Bozza
~\cite{bozzaeq} reproduce the behavior of this RI very well until
$|a|$ gets higher than $|a| \approx 0.9$ and the image moves too
far away from the equator where the ``quasi-equatorial''
approximation is no longer valid.  Choosing $a
> 0$ for a source in that position would increase $\varphi_i$ from
the beginning and would not allow us to do the comparison with the
reference.

For the magnifications, we found that our results roughly agree
with those of Bozza, except for $a \gtrsim 0.9$. However, we have
to admit that we encountered problems with the numerical
derivatives for large values of $a$. Therefore, our results for
this case should serve only as a order-of-magnitude estimate.
Nevertheless, both results agree in that the net effect of the
black hole spin is to further demagnify the RI as compared to the
Schwarzschild case ($a = 0$).

\subsection{ An Observer Off the Kerr Black Hole's Equatorial Plane ($\theta_o = \pi/4$)}

This is a more general case that will allow us to illustrate the
phenomenology of the RIs  in a more realistic astronomical
situation. We found that the general behavior of the RIs for
$\theta_o = \pi/4$ is valid for a wide range of inclinations (but
the precise location and magnifications of the images will depend
on $\theta_o$).

\begin{figure}[h]
\includegraphics[width=8.71cm,height=9.07cm]{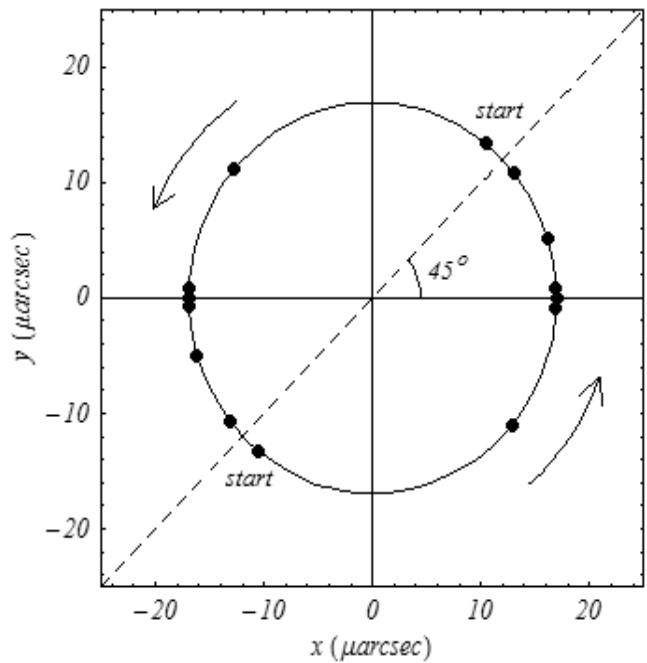}
\caption{\small Position of the lowest order RIs ($m = 3$) as a
function of the source separation. The black hole spin parameter
has been fixed to $a  = 10^{-6}$. The source is travelling along
the dashed line starting at $x_s = y_s = 1$ arcsec and finishing
at $x_s = y_s = -1$ arcsec. The arrows illustrate the direction of
movement of the two RIs, and their starting positions have been
label by the word ``start". The solid circle is the boundary of
the photon region. } \label{lowaplot}
\end{figure}

We start with a slowly rotating black hole and set $a = 10^{-6}$.
We then calculate the positions and magnifications of the two
lowest order RIs as a function of the source separation. More
precisely, we consider a source travelling along the dashed line
shown in Fig.~\ref{lowaplot}.
In the same figure, we show the positions of the two brightest
images which correspond to $m = 3,n = 1$ in all cases. The
numerical results for the two lowest order RI found are shown in
Table II for $m = 3$ and Table III for $m = 5$.
\begin{table*}
  \caption{Positions and magnifications of the lowest order RIs ($m = 3$) as a function of the source separation for
  a slowly rotating black hole. The spin parameter
  of the black hole is fixed at $a = 10^{-6}$ and the source is travelling  along the dashed line of Fig.~\ref{lowaplot}.
  Here a positive value of $\xi_s$ means that the source is in the first quadrant of the observer's sky ($\varphi_s = \pi/4$)
  and a negative value means that it is located in the third quadrant ($\varphi_s = 3\pi/4$).
  In this case we found solutions only for $n = 1$. All angular positions are
  given in microarcseconds.}
\begin{tabular}{c | c c c |c c c }
  \hline\hline
   $\xi_s$ & $x_i^+$ & $y_i^+$ & $\mu^+$ & $x_i^-$ & $y_i^-$ & $\mu^-$ \\
  \hline
$10^6$& 10.461   &13.340&$3.9\times10^{-18}$                     &$-10.461$ & $-13.340$ &$3.9\times10^{-18}$\\
$10^5$& $-12.832$& 11.078& $3.2\times10^{-17}$                   &12.832    & $-11.078$& $3.2\times10^{-17}$\\
$10^4$& $-16.932$& 0.82257& $2.4\times10^{-17}$                  &16.932    & $-0.82260$& $2.4\times10^{-17}$\\
$10^2$& $-16.952$& $7.8417\times10^{-3}$& $2.3\times10^{-17}$    &16.952    & $-7.8454\times10^{-3}$& $2.3\times10^{-17}$\\
$1$&    $-16.952$& $6.6003\times10^{-5}$ &$2.3\times10^{-17}$    &16.952    & $-6.5998\times10^{-5}$ &$2.3\times10^{-17}$\\
$-1$&   $-16.952$& $-9.1084\times10^{-5}$& $2.3\times10^{-17}$   &16.952    & $9.1081\times10^{-5}$& $2.3\times10^{-17}$\\
$-10^2$&$-16.952$& $-7.8631\times10^{-3}$& $2.3\times10^{-17}$   &16.952    & $7.8631\times10^{-3}$& $2.3\times10^{-17}$\\
$-10^4$&$-16.935$& $-0.74990$& $2.2\times10^{-17}$               &16.935    & $0.75005$& $2.2\times10^{-17}$\\
$-10^5$&$-16.161$& $-5.1170$& $1.5\times10^{-17}$                 &16.162    & $5.1154$& $1.5\times10^{-17}$\\
$-10^6$&$-13.093$& $-10.768$& $3.2\times10^{-18}$                &13.093    & $10.768$& $3.2\times10^{-18}$\\
 \hline\hline
\end{tabular}
\end{table*}
\begin{table*}
\centering
 \caption{Same as Table II but with $m = 5$. In this case we found solutions only for $n = 2$.}
\begin{tabular}{c | c c c |c c c }
  \hline\hline
   $\xi_s$ & $x_i^+$ & $y_i^+$ & $\mu^+$ & $x_i^-$ & $y_i^-$ & $\mu^-$ \\
  \hline
$10^6$& 7.9844  &14.930&$8.1\times10^{-21}$                      &$-7.9844$&$-14.930$& $8.1\times10^{-21}$  \\
$10^5$& $-16.330$& 4.4714& $2.4\times10^{-20}$                   &16.330   &$-4.4714$& $2.4\times10^{-20}$ \\
$10^4$& $-16.927$& 0.37183& $2.0\times10^{-20}$                  &16.927& $-0.37183$& $2.0\times10^{-20}$ \\
$10^2$& $-16.931$& $3.6133\times10^{-3}$& $2.0\times10^{-20}$    &16.931& $-3.6133\times10^{-3}$& $2.0\times10^{-20}$ \\
$1$&    $-16.931$& $9.3956\times10^{-6}$ &$2.0\times10^{-20}$    &16.931& $-9.3958\times10^{-6}$ &$2.0\times10^{-20}$ \\
$-1$&   $-16.931$& $-6.3395\times10^{-5}$& $2.0\times10^{-20}$   &16.931& $6.3396\times10^{-5}$& $2.0\times10^{-20}$\\
$-10^2$&$-16.931$& $-3.6657\times10^{-3}$& $2.0\times10^{-20}$   &$16.931$& $3.6657\times10^{-3}$& $2.0\times10^{-20}$\\
$-10^4$&$-16.927$& $-0.35624$& $1.9\times10^{-20}$               &16.927& 0.35624& $1.9\times10^{-20}$  \\
$-10^5$&$-16.672$& $-2.9498$& $1.6\times10^{-20}$                &16.672& 2.9498& $1.6\times10^{-20}$ \\
$-10^6$&$-13.984$& $-9.5443$& $5.2\times10^{-21}$                &13.984& $9.5443$& $5.2\times10^{-21}$\\
 \hline\hline
\end{tabular}
\end{table*}
Both images appear in opposite sides of the black hole as in the
Schwarzschild case. However, for the slowly spinning black hole,
they rotate counterclockwise as the source approaches the origin
of the observer's sky and emerges in the opposite quadrant (see
Fig.~\ref{lowaplot}). If we take $a < 0$, the effect is the same
but the images rotate clockwise. Note that for very large source
separations, the positions and magnifications of the images tend
to the familiar Schwarzschild values~\cite{ellis}.

If we compare the magnifications with those calculated in
Ref.~\cite{ellis} for a source at the same separations but with $a
= 0$ (Schwarzschild black hole), we again observe that the effect
of the black hole spin is to further demagnify the RIs.  For
instance, for a source at $\xi_s = 1 \mu$arcsec, the
magnifications calculated in \cite{ellis} for the two brightest
RIs were: $\mu_+ = \mu_- \approx 3.5\times10^{-12}$ while our
calculations for $a = 10^{-6}$ yield $\mu_+ = \mu_- \approx
2.3\times10^{-17}$, a difference of five orders of magnitude!  A
similar behavior is observed for $m = 5$ (see Table III).  This
example illustrates how even a small value of $a$ can make
significant differences in the phenomenology of the strong-field
gravitational lens.

\begin{figure}[h]
\includegraphics[width=8.71cm,height=9.07cm]{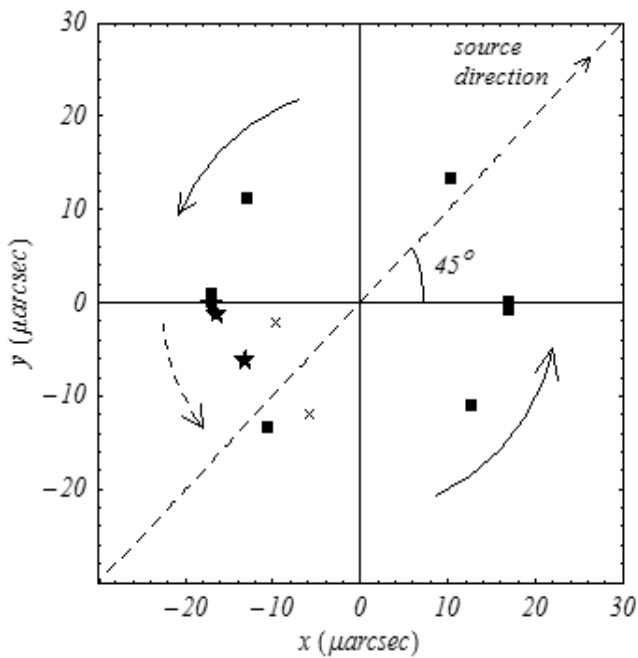}
\caption{In this plot we show the position of the lowest order RIs
as a function of the normalized angular momentum $a$. The source
is fixed at $\xi_s = 1$ arcsec, $\varphi_s = \pi/4$.  While we
increase the angular momentum starting from $a = 10^{-6}$, two
images with $m = 3$ (boxes, $\blacksquare$) move toward the $x$
axis as indicated by the solid arrows. Increasing $a$ further ($a
\gtrsim 10^{-2}$), these images disappear in the $x$ axis and give
rise to a single $m = 2$ image (stars, $\bigstar$) that move as
indicated by the dashed arrow. For $a = 0.9999$ we find two images
with $m = 2$ in the third quadrant of the observer's sky (crosses,
$\times$).} \label{largeaplot}
\end{figure}

Now we consider a fixed source at $\xi_s = 1$ arcsec, $\varphi_s =
\pi/4$ and calculate the positions and magnifications of the two
lowest order RIs as a function of the spin parameter $a$. We
consider the case of $a > 0$ only for simplicity. To illustrate
the phenomenology of the RIs we have included
Fig.~\ref{largeaplot}.
We observe that for $10^{-6} \lesssim a \lesssim 10^{-3}$ the two
lowest order sets of RIs have $m = 3$ and $m = 5$ respectively as
in the case of a Schwarzschild black hole. We also observe two
images per value of $m$.  As we increase $a$, these images move
toward the $x$ axis of the observer's sky. For $a \gtrsim 10^{-2}$
these images disappear in the $x$ axis and give rise to images of
order $m = 2$ and $m = 4$. For $m = 2$, we observe only one image
in the third quadrant of the observer's sky. On the other hand,
for $m = 4$ we found two images (one in the first quadrant and one
in the third). Increasing $a$ even more ($a = 0.9999$) we
discovered that for $m = 2$, two images are formed in the third
quadrant while for $m = 4$, a total of six images are observed:
two in the first quadrant and four in the third quadrant. The
positions of the lowest order images ($m = 2$ and $m = 3$) are
shown in Fig.~\ref{largeaplot} while the position and
magnifications of all images ($m = 2,3,4,5$) are listed in Tables
IV - VII. Is interesting how the number and order of the images
depend on the value of the spin parameter $a$.
\begin{table*}
  \caption{Positions and magnification of the order $m = 3$ RIs as a function of the black hole spin parameter $a$.
  Here we consider a fixed source at $\xi_s = 1$ arcsec and $\varphi_s = \pi/4$. For $a \gtrsim 10^{-2}$ these images
  disappear in the $x$ axis of the observer's sky. All images have $n = 1$. All angular positions are given in microarcseconds.}
\begin{tabular}{c | c c c |c c c }
  \hline\hline
   $a$ & $x_i^+$ & $y_i^+$ & $\mu^+$ & $x_i^-$ & $y_i^-$ & $\mu^-$ \\
  \hline
$10^{-6}$& 10.461  &13.340&$3.9\times10^{-18}$         &$-10.4605$&$-14.930$& $3.9\times10^{-18}$  \\
$10^{-5}$& $-12.832$& 11.077& $3.2\times10^{-18}$      &12.8318   &$-11.0777$& $3.2\times10^{-18}$ \\
$10^{-4}$& $-16.932$& 0.82137& $2.4\times10^{-19}$     &16.9326& $-0.82145$& $2.4\times10^{-19}$ \\
$10^{-3}$& $-16.947$& 0.066386& $2.3\times10^{-20}$    &16.9565& $-0.0664584$& $2.3\times10^{-20}$ \\
\hline\hline
\end{tabular}
\end{table*}
\begin{table*}
\caption{Same as Table IV but with $m = 5$. All images have $n =
2$.}
\begin{tabular}{c | c c c |c c c }
  \hline\hline
   $a$ & $x_i^+$ & $y_i^+$ & $\mu^+$ & $x_i^-$ & $y_i^-$ & $\mu^-$ \\
  \hline
$10^{-6}$& 7.9844  &14.930&$8.1\times10^{-21}$                    &$-7.9844$&$-14.930$& $8.1\times10^{-21}$  \\
$10^{-5}$& $-16.330$& 4.4712& $2.4\times10^{-21}$                  &16.330   &$-4.4713$& $2.4\times10^{-21}$ \\
$10^{-4}$& $-16.926$& 0.36919& $2.0\times10^{-22}$                &16.927& $-0.36923$& $2.0\times10^{-22}$ \\
$10^{-3}$& $-16.926$& $9.5138\times10^{-3}$& $2.0\times10^{-23}$  &16.935& $-9.5469\times10^{-3}$& $2.0\times10^{-23}$ \\
\hline\hline
\end{tabular}
\end{table*}
\begin{table*}
  \caption{Positions and magnifications of the order $m  = 2$ images as a function of the spin parameter $a$.
  Here we consider a fixed source at $\xi_s = 1$ arcsec and $\varphi_s = \pi/4$. These images appear for $a \gtrsim 10^{-2}$.
  Note that for the first three values of $a$ we have only one image  but for $a = 0.9999$ there
  are two images. Also, in this case we found images with different values of $n$ ($n = 1,2$). All
  angular positions are given in microarcseconds.}
\begin{tabular}{c | c c c c  }
  \hline\hline
   $a$ & $x_i^-$  & $y_i^-$  & $\mu^-$& $n$ \\
  \hline
$10^{-2}$&      $-16.906$  &$-0.11746$& $2.3\times10^{-21}$& 1  \\
$0.1$&          $-16.439$  &$-1.2457$& $2.5\times10^{-22}$ & 1 \\
$0.5$&          $-13.200$     & $-6.1999$& $8.3\times10^{-23}$& 1 \\
\hline $0.9999$&$-5.9311$     & $-11.907$& $1.1\times10^{-22}$& 1 \\
        &       $-9.6208$     & $-2.0417$& $1.1\times10^{-22}$& 2 \\
\hline\hline
\end{tabular}
\end{table*}
\begin{table*}
  \caption{Same as Table VI but with $m = 4$. Note that in this case we have two images for the first three values of
  $a$, but six images for $a = 0.9999$. Also, we find images with $n = 1 - 7$.}
\begin{tabular}{c | c c c c |c c c c  }
  \hline\hline
   $a$ & $x_i^+$  & $y_i^+$  & $\mu^+$ & $n$ & $x_i^-$  & $y_i^-$  & $\mu^-$& $n$  \\
  \hline
$10^{-2}$& 16.997  &0.11757&$2.3\times10^{-21}$ & 1                   &$-16.883$ &$-0.26620$& $2.0\times10^{-24}$& 2  \\
$0.1$& $17.354$& 1.2638& $2.1\times10^{-22}$& 1                  &$-16.244$   &$-2.6972$& $2.6\times10^{-25}$& 2 \\
$0.5$& $17.680$& 6.6724& $3.8\times10^{-23}$& 1 &$-7.7361$
&$-13.392$& $2.4\times10^{-25}$& 2
\\\hline
$0.9999$& $9.1375$& $15.830$& $8.1\times10^{-23}$& 1    &$-6.6670$& $-10.505$& $6.5\times10^{-25}$& 3 \\
        & 3.4194&    16.348&   $7.5 \times10^{-23}$& 1  &$-8.5479$& $-6.5516$& $8.2\times10^{-25}$& 4 \\
        &       &           &         &                 &$-9.0293$& $-4.2960$& $1.1\times10^{-24}$& 5 \\
        &       &           &         &                 &$-9.1975$& $-2.6233$& $1.7\times10^{-24}$&6 \\
        &       &           &         &                 &$-9.2611$& $-0.93129$& $6.5\times10^{-24}$&7 \\

\hline\hline
\end{tabular}
\end{table*}

With regard to the magnifications, we observe that as in previous
cases, the effect of the black hole angular momentum is to
demagnify the images (although there is a slight increase in the
magnification for $a = 0.9999$). Also note that the images with $m
= 4$ in the first quadrant have magnifications  very similar to
those with $m = 2$ (located in the third quadrant). Therefore,
these images should be considered ``dual" to each other since they
form the pair of brightest images for a given value of $a$ ($a
\gtrsim 10^{-2}$). An interesting consequence of the black hole
rotation is that for large values of $a$ ($a \gtrsim 0.1$) the two
brightest images have different magnifications. This is to be
compared  to the case of a Schwarzschild black hole where the two
brightest RIs have exactly the same magnification. Also, for large
$a$ the RIs are very static as in the Schwarzschild case (although
 for $a = 0$ they have $\varphi_i = \varphi_s$). In other words,
 their positions do not depend on the location of
the source.

 What is even more surprising  is that for $a \neq 0$ and
within our numerical precision, it seems that the ratio of the
magnifications of the two brightest RIs is insensitive to the
position of the source. This raises the possibility of extracting
information about the orientation and spin  of the black hole by
comparing the brightness of this two images. However, to move this
proposal forward we would need to study the behavior of the ratio
of their magnifications for different values of $\theta_o$. We do
not attempt to carry such analysis this time. We hope that any new
approximation scheme developed in the future can allow us to
address that question.

\section{conclusions}
In this article we have explored the phenomenology of strong field
gravitational lensing by a Kerr black hole. In particular we have
developed a general procedure to calculate the positions and
magnification of all images for an observer and source far away
from the black hole and at arbitrary inclinations. We have applied
our developed procedure to the case of a black hole at the
Galactic center with mass $M = 2.8\times 10^6\,M_{\odot}$  and at
a coordinate distance of $r_o = 8.5$ kpc. We have reproduced the
positions and magnifications of the lowest order relativistic
images found in the references for a Schwarzschild black hole and
for ``quasi-equatorial" trajectories around a Kerr black hole. We
have also presented new numerical results for the case of an
observer located at $\theta_o = \pi/4$.

Although we have not been able to give a full account of the
phenomenology for all possible combination of the source-observer
geometry and the spin parameter $a$, our limited results were
useful to get a physical insight of the effects of the black hole
angular momentum in the strong-field regime of gravitational
lensing. Moreover, our findings can serve as a ``test-bed"
calculation for any improved new model of gravitational lens.

There is no doubt that observations of the strong-field regime of
gravitational lensing will be an extremely challenging task in the
near future. This is because, as we have seen, the relativistic
images are always highly demagnified. However, if we are able to
observe them in any foreseeable future, they will provide one of
the best tests of Einstein's General Theory of Relativity in
strong gravitational fields. Moreover, as we have seen, they could
provide new tools to astrophysics by allowing the measurement of
the orientation and/or magnitude of the angular momentum of the
black hole. However, to fully confirm that this is the case, more
research is needed toward developing an analytical solution of the
strong-field gravitational lens problem in Kerr space-time.

\begin{acknowledgments}
S. E. V\'azquez is very grateful to E. P. Esteban for all his
support and advise during his undergraduate years. He would also
like to thank NSF for a Graduate Research Fellowship and the
University of California at Santa Barbara for a Broida Excellence
Fellowship. E. P. Esteban thanks the support given by UPR-Humacao
and Rice university during his sabbatical leave.
\end{acknowledgments}

\appendix*
\section{}
Here we show how to write Eqs.~(\ref{firstint}) and
(\ref{secondint}) in terms of elliptic integrals. All derivations
are done following Ref.~\cite{ellipticint}.
\subsection{Radial integrals}
We start with the radial integral of Eq.~(\ref{firstint}):

\begin{eqnarray}
\label{helliptic}
 H(\lambda ,\eta) &\equiv& 2 \int_{r_{\min}}^{\infty}
\frac{dr}{\sqrt{R(r)}} \nonumber \\
&=& \frac{4}{\sqrt{(r_a - r_c)(r_b - r_d)}} \, F(\psi,k) \;,
\end{eqnarray}
where $F(\psi,k)$  is the normal elliptic integral of the first
kind. Also,
\begin{eqnarray}
\psi = \arcsin \sqrt{\frac{r_b - r_d}{r_a - r_4}} \;, \\ \nonumber
\\ k^2 = \frac{(r_a - r_d)(r_b - r_c)}{(r_a - r_c)(r_b - r_d)}\;,
\end{eqnarray}
where $r_a = r_{\min}$, $r_b$, $r_c$, $r_d$ ($r_a > r_b > r_c >
r_d$) are the roots of $R(r) = 0$. For the radial integral of
Eq.~(\ref{secondint}) we have:

\begin{eqnarray}
\label{lelliptic}
 L(\lambda, \eta) &\equiv& 2
\int_{r_{\min}}^{\infty} \frac{a(2r -
a\lambda)}{\Delta \sqrt{R(r)}} dr \nonumber \\
&=&2a \sum_{i = 1}^2 K_i \int_{r_{\min}}^{\infty} \frac{dr}{(r -
r_i)\sqrt{R(r)}} \nonumber \\
&=& g \sum_{i = 1}^2 \Gamma_i \left[(1 - \beta_i^2)
\Pi(\psi,\alpha_i^2,k) + \beta_i^2 F(\psi,k)\right]\;, \nonumber
\\
\end{eqnarray}
where $\Pi(\psi, \alpha^2, k)$ is the normal elliptic integral of
the third kind and
\begin{eqnarray}
r_i = 1 + (-1)^{i+1}\sqrt{1 - a^2}\;, \\ \nonumber \\
 K_i = 1 +
(-1)^{i+1}\frac{1 - a\, \lambda / 2}{\sqrt{1- a^2}}\;, \\
\nonumber \\
\Gamma_i = \frac{K_i}{r_a - r_i}\;, \\ \nonumber \\
\alpha_i^2 = \frac{(r_i - r_b)(r_a - r_d)}{(r_i - r_a)(r_b -
r_d)}\;, \\ \nonumber \\
\beta_i^2 = \frac{r_i - r_a}{r_i - r_b}\;, \\ \nonumber \\
g = \frac{4 a}{\sqrt{(r_a - r_c)(r_b - r_d)}}\;.
\end{eqnarray}
These expressions are only valid for $a \neq 1$. For $a = 1$ we
have:
\begin{widetext}
\begin{eqnarray}
\label{lelliptica0}
 L(\lambda,\eta) &=& 2
\int_{r_{\min}}^{\infty} \left[ \frac{2 - \lambda}{(r - 1)^2} +
\frac{2}{r - 1} \right]
\frac{dr}{\sqrt{R(r)}} \nonumber \\
&=& \frac{g}{r_a - 1}\left\{ \beta^2 \left[ 2 + \frac{\beta^2 (2 -
\lambda)}{r_a - 1}\right]F(\psi,k) + 2(1 - \beta^2)\left[ 1 +
\frac{\beta^2 (2 - \lambda)}{r_a - 1}\right]\Pi(\psi,\alpha^2,k)
+ \frac{(1 - \beta^2)^2(2 - \lambda)}{r_a - 1} V \right\} \;, \nonumber \\
\end{eqnarray}
\end{widetext}
 where
\begin{widetext}
\begin{eqnarray}
V = \frac{1}{2(\alpha^2 - 1)(k^2 - \alpha^2)}\left[ \alpha^2 E(u)
+ (k^2 - \alpha^2) u + (2\alpha^2 k^2 + 2 \alpha^2 - \alpha^4 - 3
k^2)\Pi(\psi,\alpha^2,k) - \frac{\alpha^4 \textrm{sn}\, u\;
\textrm{cn}\, u\; \textrm{dn}\, u}{1 - \alpha^2 \textrm{sn}^2
u}\right]\;, \nonumber \\
\end{eqnarray}
\end{widetext}
\begin{eqnarray}
u = F(\psi,k)\;, \\ \nonumber \\
\alpha^2 = \frac{(1 - r_b)(r_a - r_d)}{(1 - r_a)(r_b - r_d)}\;,\\
\nonumber \\
\beta^2 = \frac{1 - r_a}{1 - r_b}\;.
\end{eqnarray}
Here $E(u)$ is the complete elliptic integral of the second kind,
and sn, cn, dn are the Jacobian elliptic functions.

\subsection{Angular integrals}
We begin by defining $u \equiv cos^2 \theta$, such that
\begin{eqnarray}
\label{difold}
 \sin \theta \, d\theta = -\frac{1}{2} \,
\textrm{sign}(\pi/2 - \theta) \frac{du}{\sqrt{u}}\;.
\end{eqnarray}
Since all angular integrals considered in this paper are symmetric
around $\theta = \pi/2$, we can restrict our interval to $\pi/2
\leq \theta \leq \pi$ and discard the sign of Eq.~(\ref{difold})
(remember that all integrals are positive definite). However, we
need to compensate for the case of a photon that crosses the
equator. To this end we define the operator
\begin{eqnarray}
\theta_1 * \theta_2 \equiv \cos\theta_1 \,\cos\theta_2\;,
\end{eqnarray}
for any two angles $\theta_1$ and $\theta_2$. Is clear that
$\theta_1 * \theta_2 > 0$ if both angles are at the same
hemisphere and negative otherwise. Therefore, is easy to show that
we can write
\begin{eqnarray}
\pm \int_{\theta_1}^{\theta_2} =
\int_{\underline{u}}^{\overline{u}} + \left[1 -
\textrm{sign}(\theta_1 * \theta_2)\right] \int_0^{\underline{u}}
\;,
\end{eqnarray}
where $\underline{u} = \min(u_1,u_2)$ and $\overline{u} =
\max(u_1,u_2)$.

In the trajectories that we consider in this article the photon
encounters various turning points defined by: $\Theta (\theta) =
0$. The positive solution to this equation is [see
Eq.~(\ref{Theta})]:

\begin{eqnarray}
u_m &=& \frac{1}{2}
\left\{1 - \frac{\lambda^2 + \eta}{a^2} \right.\nonumber \\
&&\left. + \left[\left(1 - \frac{\lambda^2 + \eta}{a^2}\right)^2 +
4\frac{\eta}{a^2} \right]^{1/2} \right\}\;,
\end{eqnarray}
where the subscript ``m" stands for ``min/max". The corresponding
angles are
\begin{eqnarray}
\theta_{\min / \max} = \arccos\left( \pm \sqrt{u_m} \right)\;,
\end{eqnarray}
where the plus sign correspond to $\theta_{\min}$ and the negative
to $\theta_{\max}$.

 Now, for a trajectory that encounters $m$ turning points
($m \geq 1$) we have
\begin{eqnarray}
\label{thetaintegrals}
 \pm \int_{\theta_s}^{\theta_{\min/ \max}} \underbrace{\pm
\int_{\theta_{\min / \max}}^{\theta_{\max/ \min}}  \pm
\int_{\theta_{\max / \min}}^{\theta_{\min/ \max}} \cdots}_{m -
1\;\; \textrm{times}} \pm
\int_{\theta_{\max / \min}}^{\theta_o} \nonumber \\
= \int_{u_s}^{u_m} + \left[1 - \textrm{sign}(\theta_s
* \theta_{ms})\right]
\int_0^{u_s} \nonumber \\
+ \int_{u_o}^{u_m} + \left[1 - \textrm{sign}(\theta_o
* \theta_{mo})\right]
\int_0^{u_o} \nonumber \\
+ 2(m - 1) \int_{0}^{u_m}\;, \nonumber \\
\end{eqnarray}
where
\begin{eqnarray}
\theta_{mo} &\equiv& \arccos\left[ \textrm{sign}( y_i) \sqrt{u_m}
\,
\right]\;, \\ \nonumber \\
\theta_{ms} &\equiv& \left\{
\begin{array}{ll}
    \theta_{mo}\; , & \hbox{ $m$ odd ;} \\
    \pi - \theta_{mo}\; , & \hbox{$m$ even ,} \\
\end{array}
\right.
\end{eqnarray}
with $y_i$ as the (possible) position of the image. In deriving
Eq.~(\ref{thetaintegrals}) we have use the fact that, as discussed
in section III, $u_m \geq u_s, u_o$. For $m = 0$ we can write
\begin{eqnarray}
\label{m=0}
 \pm \int_{\theta_s}^{\theta_s} &=&
\int_{\underline{u}}^{\overline{u}} + \left[1 -
\textrm{sign}(\theta_s * \theta_o)\right] \int_0^{\underline{u}}
\nonumber \\
&=&  \int_{\underline{u}}^{u_m} - \int_{\overline{u}}^{u_m}
 + \left[1 - \textrm{sign}(\theta_s * \theta_o)\right]
\int_0^{\underline{u}}\;, \nonumber \\
\end{eqnarray}
where $\underline{u} = \min(u_s,u_o)$ and $\overline{u} =
\max(u_s,u_o)$.

To write the angular integrals of Eq.~(\ref{firstint}) as elliptic
integrals we change variables to $u$ so that, for $\theta_j$ and
$\theta_{\min / \max}$ in the same hemisphere, we have
\begin{eqnarray}
\label{int1}
\int_{\theta_j}^{\theta_{\min / \max}}
\frac{d\theta}{\pm \sqrt{\Theta(\theta)}} &=& \frac{1}{2
|a|}\int_{u_j}^{u_m} \frac{du}{\sqrt{u(u_m - u)(u - u_3)}}
\nonumber \\ \nonumber \\
&=& h \, F(\Psi_j, \kappa)\;,
\end{eqnarray}
where
\begin{eqnarray}
h &=& \frac{1}{|a| \sqrt{u_m - u_3}} \;, \\ \nonumber \\
\label{psij}
\Psi_j &=& \arcsin \sqrt{1 - \frac{u_j}{u_m}}\;, \\ \nonumber \\
\kappa^2 &=& \frac{u_m}{u_m - u_3}\;,
\\ \nonumber \\
 u_3 &=& \frac{1}{2}
\left\{1 - \frac{\lambda^2 + \eta}{a^2} \right.\nonumber \\
&&\left. - \left[\left(1 - \frac{\lambda^2 + \eta}{a^2}\right)^2 +
4\frac{\eta}{a^2} \right]^{1/2} \right\}\;,
\end{eqnarray}
and $u_m > u_3$ ($u_3 < 0$). On the other hand we have
\begin{eqnarray}
\label{int2}
 \frac{1}{2 |a|} \int_0^{u_j}\frac{du}{\sqrt{u(u_m -
u)(u - u_3)}} =  h \, F(\Phi_j, \kappa)\;,
\end{eqnarray}
where
\begin{eqnarray}
\label{phij} \Phi_j = \arcsin \sqrt{\frac{u_j (u_m - u_3)}{u_m(u_j
- u_3)}}\;.
\end{eqnarray}

Using Eqs.~(\ref{thetaintegrals}), (\ref{m=0}), (\ref{int1}) and
(\ref{int2}) we can write the left hand side of
Eq.~(\ref{firstint}) as
\begin{eqnarray}
G(\lambda, \eta, \theta_s, m) &\equiv& h \left\{F(\Psi_s,\kappa) +
F(\Psi_o,\kappa) \right. \nonumber \\
&& \left.+ \left[1 - \textrm{sign}(\theta_s
* \theta_{ms})\right] F(\Phi_s,\kappa)\right.\nonumber \\
&& \left. + \left[1 - \textrm{sign}(\theta_o * \theta_{mo})\right]
F(\Phi_o,\kappa) \right. \nonumber \\
&&\left. + 2(m - 1)K(\kappa) \right\}\;,
\end{eqnarray}
for $m \geq 1$. Here, $K(\kappa)$ is the compete elliptic integral
of the first kind: $K(\kappa) = F(\pi/2,\kappa)$. For $m = 0$ we
have
\begin{eqnarray}
G(\lambda, \eta, \theta_s, m) &\equiv& h
\left\{F(\underline{\Psi},\kappa)
-F(\overline{\Psi},\kappa) \right. \nonumber \\
&& \left.+ \left[1 - \textrm{sign}(\theta_s
* \theta_{o})\right] F(\underline{\Phi},\kappa)
\right\}\;,\nonumber\\
\end{eqnarray}
where $\underline{\Psi}$ and  $\overline{\Psi}$ are given by
Eq.~(\ref{psij}) with the substitutions $u_j \rightarrow
\min(u_s,u_o)$ and  $u_j \rightarrow \max(u_s,u_o)$ respectively,
and $\underline{\Phi}$ is given by Eq.~(\ref{phij}) with the
substitution $u_j \rightarrow \min(u_s,u_o)$.

Now we turn our attention to the angular integrals of
Eq.~(\ref{secondint}). Making the usual change of variable, we
get:
\begin{eqnarray}
\label{int1j}
 &&\int_{\theta_j}^{\theta_{\min / \max}}
\frac{d\theta}{\pm \sin^2
\theta \sqrt{ \Theta(\theta)}} \nonumber \\
&&= \frac{1}{2 |a|}\int_{u_j}^{u_m} \frac{du}{(1 - u)\sqrt{u(u_m -
u)(u - u_3)}}
\nonumber \\ \nonumber \\
&&= \frac{h}{1 - u_m}\Pi(\Psi_j,\gamma^2,\kappa)\;,
\end{eqnarray}
where
\begin{eqnarray}
\gamma^2 = \frac{u_m}{u_m - 1}\;.
\end{eqnarray}
The second integral we need is
\begin{eqnarray}
\label{int2j}
 &&\frac{1}{2 |a|}\int_{0}^{u_j} \frac{du}{(1 -
u)\sqrt{u(u_m -
u)(u - u_3)}} \nonumber \\
&&= \frac{h}{1 - u_3} \left[ F(\Phi_j,\kappa) - u_3
\Pi(\Phi_j,\rho^2,\kappa) \right] \;, \nonumber \\
\end{eqnarray}
where
\begin{eqnarray}
\rho^2 = \frac{u_m(1 - u_3)}{u_m - u_3}\;.
\end{eqnarray}

Using (\ref{thetaintegrals}), (\ref{m=0}), (\ref{int1j}) and
(\ref{int2j}) the angular integrals of Eq.~(\ref{secondint})
become

\begin{widetext}
\begin{eqnarray}
\label{Jelliptic}
 J(\lambda, \eta, \theta_s, m) &\equiv&
\frac{h\,\lambda}{1-u_m}\left[\Pi(\Psi_s,\gamma^2,\kappa) +
\Pi(\Psi_o,\gamma^2,\kappa)\right] + \frac{h\,\lambda}{1 - u_3}
\left\{\left(1 - \textrm{sign}(\theta_s
* \theta_{ms}) \right)\left[ F(\Phi_s,\kappa) - u_3
\Pi(\Phi_s,\rho^2,\kappa) \right] \right. \nonumber \\
&& \left. + \left(1 - \textrm{sign}(\theta_o
* \theta_{mo}) \right)\left[ F(\Phi_o,\kappa) - u_3
\Pi(\Phi_o,\rho^2,\kappa) \right] + 2(m-1)\left[ K(\kappa) - u_3
\Pi(\rho^2,\kappa) \right] \right\}\;,
\end{eqnarray}
\end{widetext}
for $m \geq 1$, and
\begin{widetext}
\begin{eqnarray}
\label{Jelliptic2}
 J(\lambda, \eta, \theta_s, m) &\equiv&
\frac{h\,\lambda}{1-u_m}\left[
\Pi(\underline{\Psi},\gamma^2,\kappa) -
\Pi(\overline{\Psi},\gamma^2,\kappa)\right] + \frac{h\,\lambda}{1
- u_3} \left\{\left[1 - \textrm{sign}(\theta_s
* \theta_{o}) \right]\left[ F(\underline{\Phi},\kappa) - u_3
\Pi(\underline{\Phi},\rho^2,\kappa) \right] \right\}\;, \nonumber
\\
\end{eqnarray}
\end{widetext}
for $m = 0$.

For the case of a Schwarzschild black hole ($a = 0$) the angular
integrals can be solved in close form since the polynomial in the
square root become one of second order:
\begin{eqnarray}
&&\int_{\theta_j}^{\theta_{\min / \max}} \frac{d\theta}{\pm
\sqrt{\Theta(\theta)}}\nonumber \\
 &&= \frac{1}{2}
\int_{u_j}^{u_m}
\frac{du}{\sqrt{\eta\, u - (\lambda^2 + \eta)u^2}} \nonumber \\
&&= \frac{u_m}{2\sqrt{\eta}} \left[ \frac{\pi}{2} + \arcsin\left(1
- 2\frac{u_j}{u_m}\right)\right]\;,
\end{eqnarray}
where
\begin{eqnarray}
u_m = \frac{\eta}{\lambda^2 + \eta}\;.
\end{eqnarray}
Similarly,
\begin{eqnarray}
&&\frac{1}{2} \int_{0}^{u_j}
\frac{du}{\sqrt{\eta\, u - (\lambda^2 + \eta)u^2}} \nonumber \\
&&= \frac{u_m}{2\sqrt{\eta}} \left[ \frac{\pi}{2} - \arcsin\left(1
- 2\frac{u_j}{u_m}\right)\right]\;.
\end{eqnarray}
Therefore, using Eqs.~(\ref{thetaintegrals}) and (\ref{m=0}), the
right hand side of Eq.~(\ref{firstint}) becomes
\begin{eqnarray}
G(\lambda,\eta,\theta_s, m) &\equiv& \frac{u_m}{2\sqrt{\eta}}
\left[
2\pi m  \right. \nonumber \\
&& \left. + \textrm{sign}(\theta_s
* \theta_{ms}) \arcsin \left(1 - 2\frac{u_s}{u_m}\right)\right. \nonumber
\\
 && \left. +   \textrm{sign}(\theta_o
* \theta_{mo}) \arcsin \left(1 - 2\frac{u_o}{u_m} \right)
\right]\;,\nonumber \\
\end{eqnarray}
for $m \geq 1$, and
\begin{eqnarray}
G(\lambda,\eta,\theta_s, m) &\equiv&
\frac{u_m}{2\sqrt{\eta}}\left\{\frac{\pi}{2} - \arcsin \left(1 -
2\frac{\overline{u}}{u_m}\right) \right. \nonumber \\
&& \left. - \textrm{sign}(\theta_s
* \theta_o) \left[\frac{\pi}{2} - \arcsin \left(1 -
2\frac{\underline{u}}{u_m}\right) \right] \right\}\;, \nonumber \\
\end{eqnarray}
for $m = 0$.

The angular integrals of Eq.~(\ref{secondint}) can also be
calculated in close form for $a = 0$. For that, is convenient to
introduce a new variable: $w \equiv \sin^2 \theta = 1 - u$. Thus,
\begin{eqnarray}
 &&\int_{\theta_j}^{\theta_{\min / \max}}
\frac{d\theta}{\pm \sin^2
\theta \sqrt{ \Theta(\theta)}} \nonumber \\
&&= \frac{1}{2}\int_{w_m}^{w_j} \frac{dw}{w\sqrt{-\lambda^2 +
(2\lambda^2 + \eta) w - (\lambda^2 + \eta) w^2}} \nonumber \\
&&= \frac{1}{2|\lambda|} \left[ \frac{\pi}{2} + \arcsin\left(1 -
\frac{2 \lambda^2 u_j^2}{\eta(1 - u_j)} \right)\right]\;.
\end{eqnarray}

Similarly, for the $0 \rightarrow u_j$ integration  we have
\begin{eqnarray}
&&\frac{1}{2}\int_{w_j}^{1} \frac{dw}{w\sqrt{-\lambda^2 +
(2\lambda^2 + \eta) w - (\lambda^2 + \eta) w^2}}\nonumber \\
 &&=\frac{1}{2|\lambda|} \left[ \frac{\pi}{2} - \arcsin\left(1 -
\frac{2 \lambda^2 u_j^2}{\eta(1 - u_j)} \right)\right]\;.
\end{eqnarray}
Therefore, the angular integrals of Eq.~(\ref{secondint}) become
\begin{widetext}
\begin{eqnarray}
J(\lambda,\eta,\theta_s, m) &\equiv&
\frac{\textrm{sign}(\lambda)}{2} \left[ 2\pi m +
\textrm{sign}(\theta_s
* \theta_{ms}) \arcsin\left(1 -
\frac{2 \lambda^2 u_s^2}{\eta(1 - u_s)} \right) +
\textrm{sign}(\theta_o
* \theta_{mo}) \arcsin\left(1 -
\frac{2 \lambda^2 u_o^2}{\eta(1 - u_o)} \right)
\right]\;,\nonumber \\
\end{eqnarray}
\end{widetext}
for $m \geq 1$, and
\begin{widetext}
\begin{eqnarray}
J(\lambda,\eta,\theta_s, m) &\equiv&
\frac{u_m}{2\sqrt{\eta}}\left\{\frac{\pi}{2} - \arcsin\left(1 -
\frac{2 \lambda^2 \overline{u}^2}{\eta(1 - u_o)} \right)  -
\textrm{sign}(\theta_s
* \theta_o) \left[\frac{\pi}{2} - \arcsin\left(1 -
\frac{2 \lambda^2 \underline{u}^2}{\eta(1 - u_o)} \right) \right]
\right\}\;,
\end{eqnarray}
\end{widetext}
for $m = 0$.

Using the expressions for $H$, $G$, $L$ and $J$ given above, the
``lens equations" (\ref{firstint}) and (\ref{secondint}) become:
\begin{eqnarray}
H(\lambda, \eta) - G(\lambda, \eta, \theta_s, m) &=& 0 \;, \\
\nonumber
\\
\Delta\phi - L(\lambda, \eta) - J(\lambda, \eta, \theta_s, m) &=&
0\;,
\end{eqnarray}
with
\begin{eqnarray}
 \Delta\phi = \left\{
\begin{array}{ll}
    -\phi_s - 2\pi n &, \;{L + J < 0} \\
    2\pi( n +1) - \phi_s  &,\; {L + J > 0} \\
\end{array}
\right.
\end{eqnarray}
where $n = 0,1,2 \ldots$ is the number of windings around the $z$
axis, and $\lambda$, $\eta$ and $\theta_s$ can be written in terms
the observer's sky coordinates by using Eqs.~(\ref{lambda}),
(\ref{eta}), (\ref{xs}) and (\ref{ys}).

\end{document}